\documentclass[useAMS,usenatbib,usegraphicx]{mn2e}

\usepackage[dvips,dvipdfm]{graphicx,color}
\usepackage{aas_macros}
\title[Dry minor mergers and size evolution of ETGs]{Dry minor mergers and size evolution of high-z compact massive early-type galaxies}
\author[T. Oogi and A. Habe]{Taira Oogi$^{1}$\thanks{E-mail: oogi@astro1.sci.hokudai.ac.jp (TO)} and Asao Habe$^{1}$\thanks{E-mail: habe@astro1.sci.hokudai.ac.jp (AH)} \\
$^{1}$Department of Cosmosciences, Graduate School of Science, Hokkaido University, Sapporo, Japan}

\begin{document}

\date{}

\pagerange{\pageref{firstpage}--\pageref{lastpage}} \pubyear{2011}

\maketitle

\label{firstpage}

\begin{abstract}
Recent observations show evidence that high-z ($z\sim 2 - 3$) early-type galaxies (ETGs) are more compact than those with comparable mass at $z\sim 0$. Such a size evolution is most likely explained by the `Dry Merger Sceanario'. However, previous studies based on this scenario are not able to consistantly explain both the properties of the high-z compact massive ETGs and the local ETGs. We investigate the effect of multiple sequential dry minor mergers on the size evolution of the compact massive ETGs.
From an analysis of the Millennium Simulation Database, we show that such minor (stellar mass ratio $M_{2}/M_{1} < 1/4$) mergers are extremely common during hierarchical structure formation.
We perform N-body simulations of sequential minor mergers with parabolic and head-on orbits, including a dark matter component and a stellar component.
Typical mass ratios of the minor mergers are $1/20 < M_{2}/M_{1} \lid 1/10$.
We show that sequential minor mergers of compact satellite galaxies are the most efficient at promoting size growth and decreasing the velocity dispersion of the compact massive ETGs in our simulations.
The change of stellar size and density of the merger remnants is consistent with recent observations.
Furthermore, we construct the merger histories of candidates for the high-z compact massive ETGs using the Millennium Simulation Database, and estimate the size growth of the galaxies by the dry minor merger scenario.
We can reproduce the mean size growth factor between $z=2$ and $z=0$, assuming the most efficient size growth obtained during sequential minor mergers in our simulations.
However, we note that our numerical result is only valid for merger histories with typical mass ratios between 1/20 and 1/10 with parabolic and head-on orbits, and that our most efficient size growth efficiency is likely to an upper limit.
\end{abstract}


\section{Introduction}

There is significant observational evidence that massive ($\ga 10^{11} M_{\odot}$) early-type galaxies (hereafter ETGs) at high-z ($z\sim 2 - 3$) are more compact than galaxies with comparable mass in the local Universe.
However, a physical explanation of the size evolution from such compact ETGs to the local ETGs is still lacking within the $\Lambda$CDM galaxy formation framework.

Many observations of the high-z ETGs show that the average size of the high-z massive ETGs is smaller by a factor of three to five than that of the local ETGs with comparable mass (e.g. \citealt{2006ApJ...650...18T}, \citealt{2007MNRAS.382..109T}, \citealt{2008ApJ...687L..61B}, \citealt{2008A&A...482...21C}, \citealt{2008ApJ...677L...5V}). These high-z ETGs also lie below the stellar mass-size relation in the local Universe; a relation well investigated through the Sloan Digital Sky Survey (SDSS) database (e.g. \citealt{2003MNRAS.343..978S}).
Additionally, observed high-z ETGs indicate that stellar densities of the high-z ETGs are $1-2$ order of magnitude higher than the local ETGs. Several studies claim that this size growth is even more significant for more massive ETGs (e.g. \citealt{2010ApJ...717L.103N}). Recently, several observations have measured the velocity dispersions of the high-z ETGs (e.g. \citealt*{2009Natur.460..717V}).
The observed velocity dispersions are also higher than the average velocity dispersion of the local ETGs with comparable mass (\citealt{2009ApJ...696L..43C}, \citealt{2009ApJ...704L..34C}, \citealt{2009Natur.460..717V}, \citealt{2011ApJ...736L...9V}).
The compact massive ETGs at high-z are already quiescent in star formation activity and have an old stellar population (e.g. \citealt{2008A&A...482...21C}, \citealt{2008ApJ...677L...5V}) similar to ETGs in the local Universe (e.g. \citealt{2005ApJ...621..673T}).

Such compact massive ETGs are extremely rare in the local Universe.
\citet{2010ApJ...720..723T} have searched for compact ETGs in the local Universe with the SDSS database but have only found less massive compact ETGs which are not readily comparable to the ETGs found at high-z. \citet{2008A&A...482...21C} also found that such superdense ETGs are extremely rare at $z \approx 0$.
On the other hand, \citet{2010ApJ...712..226V} claim to find a small number of compact massive ETGs in the nearby X-ray-selected clusters and claim that these galaxy's properties correspond to the high-z compact massive ETGs.
This last example notwithstanding, it is clear observations indicate that most of the compact massive ETGs at high-z must increase their sizes and must decrease their velocity dispersion from $z\simeq 2$ to $z=0$ without recent star formation, in order to keep their old stellar populations.
This problem is called as ``size evolution problem of ETGs''.

Dry (i.e., gas-poor, collisionless) mergers are a popular scenario for the size evolution of the ETGs.
Dry mergers can increase the galaxy's size more than wet (i.e., gas-rich) mergers because of the absence of energy dissipation via the gas component (e.g. \citealt{2006ApJ...650..791C}, \citealt*{2007ApJ...658...65C}).
Furthermore, dry mergers keep the old stellar populations of progenitor galaxies without the addition of new star formation.
There is direct observational evidence for dry mergers (e.g. \citealt{2006ApJ...640..241B}).
Using a semi-analytic approach to galaxy formation, \citet{2006ApJ...648L..21K} show that massive ETGs experience dry mergers at low redshift ($z<2$) and suggest that dry mergers are the main driver for the size evolution of ETGs.
There are alternative scenarios to dry mergers that involve expansion of the stellar system due to a significant mass loss, either by quasars (\citealt{2008ApJ...689L.101F}, \citealt{2011MNRAS.414.3690R}) or stellar winds (\citealt{2009ApJ...695..101D}).
However, the amount of the stellar mass loss is too small to explain the size evolution (\citealt{2009ApJ...695..101D}), the compact massive ETGs already are quiescent in star formation activity, and they do not seem to have sufficient gas to puff up the stellar system (\citealt{2009ApJ...697.1290B}).

Dry minor mergers are a more effective process for the size evolution of ETGs than dry major mergers.
Using the virial theorem, \citet{2009ApJ...697.1290B} and \citet*{2009ApJ...699L.178N} make a simple estimate of the size growth for dry mergers.
They use initial galaxy models that consist purely of stellar systems and only assume energy conservation during mergers on parabolic orbits.
They obtain that for dry minor mergers, the galaxy sizes increase by the square of the mass increase rather than linearly for equal-mass mergers.
Semi-analytic models of hierarchical galaxy formation show that minor mergers are an important channel of mass assembly for massive galaxies in $z\la1-2$ (e.g. \citealt{2006MNRAS.370..902K}, \citealt{2008MNRAS.384....2G}).
Observational studies suggest that even for massive galaxies the major merger rate is low (e.g. \citealt{2009ApJ...697.1369B}).
It is not likely that the major mergers explain the extent of the size evolution of ETGs (\citealt{2010ApJ...720..723T}).
Therefore, dry minor mergers are likely to be the main driver for the size evolution of the massive ETGs.

While a number of numerical simulations have been done to study the size evolution problem by dry mergers (e.g. \citealt{2009ApJ...691.1424H}, \citealt{2009ApJ...699L.178N}, \citealt*{2009ApJ...703.1531N}, \citealt{2009ApJ...706L..86N}, \citealt{2010MNRAS.401.1099H}),
they remain insufficient to explain the observed difference between high z and local ETGs.
Several studies have performed high-resolution N-body simulations of dry major and minor mergers to investigate the size evolution problem (\citealt{2009ApJ...703.1531N}, \citealt{2009ApJ...706L..86N}, \citealt{2012MNRAS.tmp.2680N}, \citealt{2012arXiv1206.1597H}).
\citet{2009ApJ...706L..86N} simulate a set of dry major and minor mergers and
examine consistency of their results with the stellar-mass scaling relations of the local ETGs obtained from the Sloan Lens ACS (SLACS) Survey (\citealt{2008ApJ...682..964B}).
They conclude that typical local massive ETGs can not have assembled more than $\sim45$ \% of their stellar mass nor grown more than a factor of $\sim1.9$ in size via dry mergers.
While they did explore the effect of minor mergers, they limited their study to only the cases of ``multiple'' and ``simultaneous'' minor mergers.
However, theoretical studies on merger rates indicate that simultaneous minor mergers are extremely rare (e.g. \citealt{2008MNRAS.384....2G}, \citealt{2009ApJ...702.1005S}, \citealt{2010ApJ...715..202H}).
In the hierarchical structure formation scenario in the $\Lambda$CDM cosmology, ``sequential'' minor mergers are more common than simultaneous events.

In this paper, we investigate the effects of dry major and minor mergers on the size evolution of the ETGs with N-body simulations, particularly focusing on effects of the sequential dry minor mergers and of the effect of compactness of the satellite galaxies on the size evolution.
To explore the effects of the mergers in inner regions of the compact massive ETGs, we take care to simulate dry minor mergers with sufficient spacial and time resolution.
We show that the sequential minor mergers of the compact satellite galaxies lead to the most efficient size growth and the most efficient decrease of velocity dispersion.
Furthermore, we construct the merger histories of candidates of the high-z compact massive ETGs using the Millennium Simulation Database --a public galaxy catalog with a semi-analytic model based on the $\Lambda$CDM cosmology (\citealt{2005Natur.435..629S}, \citealt{2007MNRAS.375....2D})-- and use this to model size growth of galaxies from dry minor mergers.
Based on the size growth efficiency derived by our N-body simulations, we discuss the size growth of galaxies from $z\sim2$ to $z=0$.

The paper is organized as follows.
In \S 2, we show the importance of minor mergers for evolution of the high-z massive (stellar mass : $M_{*} \gid 10^{11}M_{\odot}$) galaxies using the Millennium Simulation Database.
In \S 3, we describe the outline of the simulations, presenting our initial models and choice of simulation parameters in more detail.
Results of the analysis of the merger remnants of our simulations are presented in \S 4.
In \S 5, we compare our results with observations, and discuss the size growth of the compact massive ETGs.
In \S 6, we give a summary of the paper.

\section{Cosmological merger histories of massive galaxies}

\subsection{Sample selection}
To understand the evolution of the ETGs, we examine mass assembly histories of the high-z compact massive ETGs.
We get merging history information of the dark matter haloes and those of the stellar bulges which formed in centres of the dark matter haloes, from the results of a semi-analytic model in the Millennium Simulation Database (\citealt{2005Natur.435..629S}, \citealt{2007MNRAS.375....2D}).
While \citet{2007MNRAS.375....2D} use strong supernova and active galactic nucleus (AGN) feedback models to reproduce observed properties of the brightest cluster galaxies (BCGs), they also adjust their model parameters in order to be in good agreement with the observations of the local Universe (see also \citet{2006MNRAS.365...11C} using a similar model).
Thus, we use this database as a galaxy formation model for all galaxies.
We extract galaxy samples at $z=0$ whose main progenitors have a bulge mass more than $10^{11} M_{\odot}$ at $z=2.07$.
These galaxies are candidates for the high-z compact massive ETGs (\citealt{2008ApJ...677L...5V}).
We found 1045 galaxies satisfying this condition in the Millennium Simulation Database.
In the following subsections, we analyse the merger histories of these sampled galaxies.

\subsection{The important role of minor mergers in galaxy mass growth}
\begin{figure*}
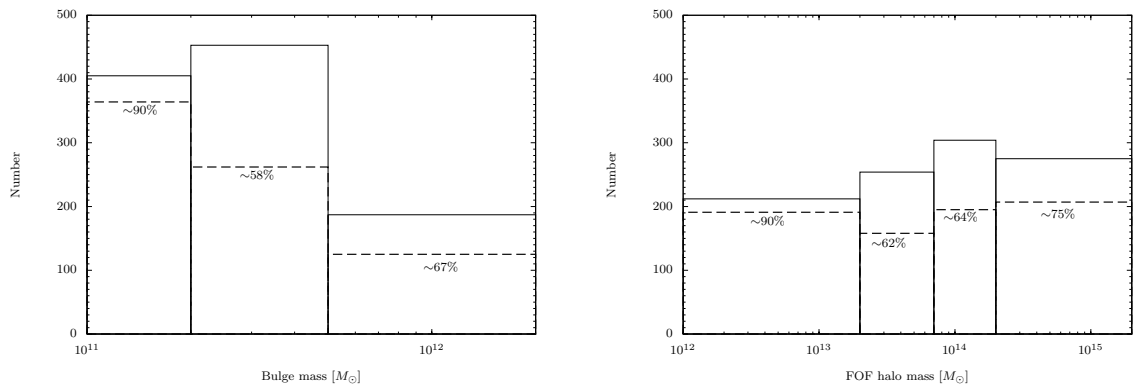

  \begin{minipage}{180mm}
    \begin{center}
      \begin{tabular}{cc}
        \resizebox{75mm}{!}{\input{./picture/number_of_dominant120419.tex}} &
        \resizebox{75mm}{!}{\input{./picture/z0dm_number_of_dominant120419.tex}} \\
      \end{tabular}
      \caption{
        Left: Number of our sample galaxies derived by the Millennium Simulation Database (solid lines).
        Horizontal axis shows the bulge mass of the galaxies at $z=0$.
        Vertical axis shows the number of galaxies in each mass bin.
        The mass bins are [$1.0\times 10^{11}:2.0\times10^{11}$], [$2.0\times 10^{11}:5.0\times10^{11}$], and [$5.0\times 10^{11}:2.0\times10^{12}$] in units of $M_{\odot}$.
        Fraction of minor merger-dominated galaxies (see text) are also plotted (dashed lines with percentages).
        Right: The same as the left panel, but for the horizontal axis that shows the FOF halo mass of the galaxies at $z=0$.
        The mass bins are [$1.0\times 10^{12}:2.0\times10^{13}$], [$2.0\times 10^{13}:7.0\times10^{13}$], [$7.0\times 10^{13}:2.0\times10^{14}$], and [$2.0\times 10^{14}:2.0\times10^{15}$] in units of $M_{\odot}$.
      }
      \label{number_of_dominant}
    \end{center}
  \end{minipage}
\end{figure*}

Fig. \ref{number_of_dominant} shows the number of the sample galaxies at different stellar mass taken at $z=0$ (solid lines).
These galaxies have experienced major and minor mergers from $z=2.07$ to $z=0$.
We define the minor mergers as mergers with mass ratios $M_{2}/M_{1} < 1/4$, otherwise we define them as major mergers.
We also define `minor merger-dominated galaxies' as galaxies for which the minor mergers are the dominant form of mass increase during the period $z=2.07$ to $z=0$, and we define `major merger-dominated galaxies' as galaxies for which the major mergers are the dominant effect.
In the left panel of Fig. \ref{number_of_dominant}, we show the fraction of the minor merger-dominated galaxies in each bulge mass bin, while in the right panel of Fig. \ref{number_of_dominant}, we show the fraction in each mass bin for the main dark matter haloes within which the galaxies reside (we call these the ``FOF'' halo).
We find that $\sim 65\%$ of the total sample galaxies are the minor merger-dominated galaxies, and $\sim 38 \%$ of the total sample galaxies have never experienced major mergers from $z=2.07$ to $z=0$. 

\begin{figure*}
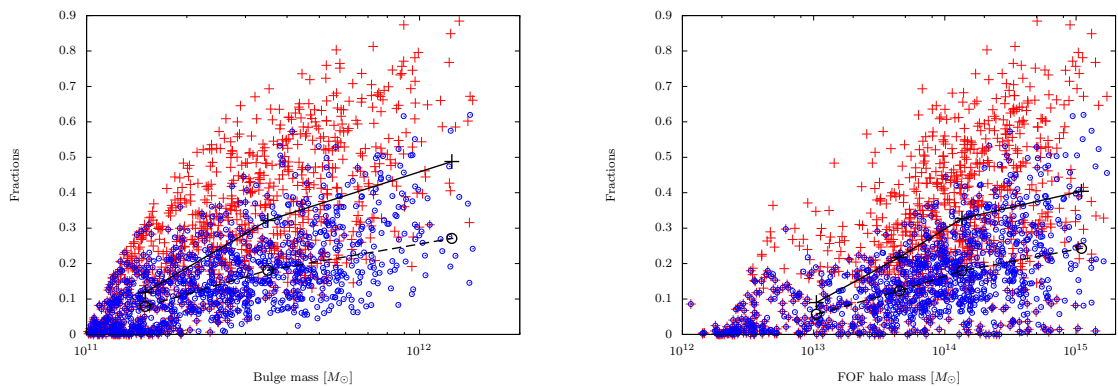

    \begin{minipage}{180mm}
        \begin{center}
            \begin{tabular}{cc}
                \hspace{-8mm}
                \resizebox{75mm}{!}{\input{./picture/z0bulge_mean_frac20120721.tex}} &
                \resizebox{75mm}{!}{\input{./picture/z0dm_mean_frac20120721.tex}}
            \end{tabular}
            \caption{
            Left: Fractional cumulative mass growth by minor mergers with mass ratios $M_{2}/M_{1} < 1/4$ and $M_{2}/M_{1} < 1/10$ from $z=2.07$ to $z=0$ for our sample galaxies derived by the Millennium Simulation Database.
            Here, the fractional cumulative mass growth is the fraction of mass increased by the minor mergers from $z=2.07$ to $z=0$ to the mass of the bulge at $z=0$.
            Horizontal axis shows the bulge mass at $z=0$.
            Red crosses represent the fractional cumulative mass growth with mass ratios $M_{2}/M_{1} < 1/4$.
            We average these in the same mass bins of Fig. \ref{number_of_dominant} and show the average values with black crosses and the solid line.
            Blue open circles represent that with mass ratios $M_{2}/M_{1} < 1/10$.
            We also average these and show them with black open circles and the dashed line.
            Right : 
            The same as the left panel, but for the horizontal axis that shows the FOF halo mass of the galaxies at $z=0$.
            }
            \label{minor_dominated_fraction}
        \end{center}
    \end{minipage}
\end{figure*}

In the left panel of Fig. \ref{minor_dominated_fraction}, we show that fractional cumulative mass growth of the sample galaxies by minor mergers from $z=2.07$ to $z=0$.
Here, the fractional cumulative mass growth is the fraction of mass increased by minor mergers from $z=2.07$ to $z=0$ to the bulge mass at $z=0$.
We find that on average the fractional cumulative mass growth by the minor mergers with mass ratios $M_{2}/M_{1} < 1/4$ ($M_{2}/M_{1} < 1/10$) is $\sim 0.3$ ($\sim 0.2$), and more massive galaxies have larger fractional cumulative mass growth by the minor mergers.
Thus, minor mergers are one of the dominant processes for the mass growth of our sample galaxies.
In the right panel of Fig. \ref{minor_dominated_fraction}, we show the dependence of the fractional cumulative mass growth on the mass of the FOF haloes.
We find that many galaxies have increased their mass significantly by the minor mergers in more massive FOF haloes.
Therefore, we expect that these minor mergers increase galaxy sizes, in particular, in high density environments.
ETGs are likely to be in such high density environments, which is well known as the morphology-density relation (\citealt{1980ApJ...236..351D}).
For this reason, in this paper we concentrate the minor mergers as the one of the important process driving the size evolution of ETGs.

\subsection{Stellar merger histories of sample galaxies}
We analyze the merging events of the 1045 sample galaxies, and show that the minor merger events are sequential processes rather than simultaneous events between $z=2.07$ and $z=0$.
We obtain merging epochs at which a satellite (less massive) galaxy merges into the primary (more massive) galaxy from the results of the semi-analytic model (see \citealt{2006MNRAS.365...11C}).
We identify 19605 minor mergers, where we count multiple minor mergers as one minor merging event if they occur in the time interval of $\sim 0.35$ Gyr, corresponding to the dynamical time-scale of the dark matter haloes of the galaxies at $z\sim2$.
Except for one case, the sum of the accreted mass through the minor mergers  during this time interval is less than $50\%$ of the mass of the main progenitor galaxy.
Therefore, almost all minor mergers occur sequentially.

\begin{figure*}
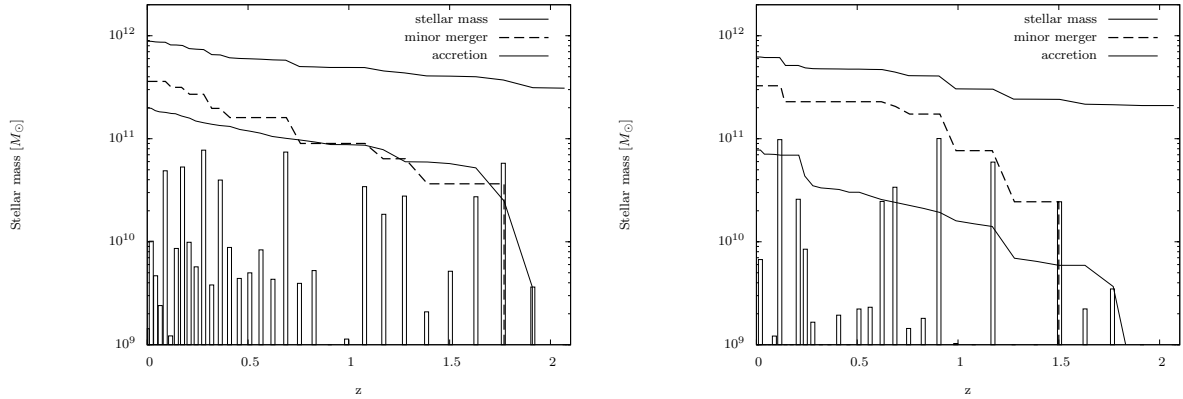

    \begin{minipage}{180mm}
        \begin{center}
            \begin{tabular}{cc}
                \hspace{-8mm}
                \resizebox{80mm}{!}{\input{./picture/436017494000000_120419.tex}} 
                \resizebox{80mm}{!}{\input{./picture/140005829362513_120419.tex}} 
            \end{tabular}
            \caption{
            Stellar merger histories of our two sample galaxies derived by the Millennium Simulation Database
            These have never experienced major mergers.
            Left: a central galaxy of a FOF halo which mass is $1.5 \times 10^{15} M_{\odot}$.
            Right:  a member galaxy of a FOF halo which mass is $1.8 \times 10^{14} M_{\odot}$.
            Thick solid line is stellar mass of the galaxies as a function of redshift.
            Bars in each panel show increased mass by minor mergers at each event.
            Dashed and dotted lines show cumulative increased stellar mass from $z=2.07$ to $z=0$ as minor mergers and accretions, respectively.
            We define accretions as mergers with mass ratios less than 1:20.
            }
            \label{merger_history}
        \end{center}
    \end{minipage}
\end{figure*}

Fig. \ref{merger_history} shows the stellar merger histories of two sample galaxies.
One is a central galaxy of a FOF halo which mass is $1.5 \times 10^{15} M_{\odot}$, and another is a member galaxy of a FOF halo which mass is $1.8 \times 10^{14} M_{\odot}$.
These two galaxies have experienced sequential minor mergers, but have not experienced major mergers since $z=2.07$.
We find that the total merged mass through minor mergers is comparable to the stellar mass at $z=2.07$.
Therefore, the size evolution is driven significantly by this process.
In this paper, we focus such a ``sequential'' minor merging processes.

\section{N-body simulations}\label{sec:nbodysimulation}
In this section, we describe our initial galaxy models, model parameters, N-body simulation methods, and analyzing methods of merger remnants.

\subsection{Initial galaxy models}\label{ssec:initial}

Our model galaxies consist of two components: a stellar bulge and a dark matter halo.
We model the dark matter mass distribution with a \citet{1990ApJ...356..359H} profile which has the total mass, $M_{dm}$:

\begin{equation}
  \rho_{dm} (r) = \frac {M_{dm}} {2\pi} \frac {a_{dm}} {r(r+a_{dm})^3},
\end{equation}
scaled to match the NFW profile found in cosmological simulations \citep*{1997ApJ...490..493N},
as described in \citet*{2005MNRAS.361..776S}.
In this model, the total mass, $M_{dm}$, corresponds to the mass within the virial radius of the NFW model and the inner density profile of the Hernquist profile is also equal to that of the NFW model.
Satisfying the above condition, the scale radius of the Hernquist profile $a_{dm}$, is related to the scale length $r_s$ of the NFW halo and halo concentration $c_{vir} (c_{vir}=r_{vir}/r_{s})$ as

\begin{equation}
  a_{dm} = r_s \sqrt{2[\ln(1+c_{vir})-c_{vir}/(1+c_{vir})]}.
\end{equation}
We derive the virial radius of the halo from a spherical collapse model, $r_{vir} = r_{vir} (M_{dm}, z)$ (e.g. \citealt{2001PhR...349..125B}), where $z$ is the formation redshift of the halo.
Using these parameters of mass, redshift and halo concentration for a dark matter halo, we can obtain the corresponding halo model for N-body realization.

We also model the initial distribution of the bulge with a Hernquist profile,
\begin{equation}
  \label{eq:hern_star}
  \rho_{*} (r) = \frac {M_{*}} {2\pi} \frac {a} {r(r+a)^3},
\end{equation}
with total stellar mass $M_{*}$, quarter-mass radius $r_{1/4}=a$ and half-mass radius $r_{1/2}=(1+\sqrt{2})a$.
Observationally, a galaxy size is usually characterized by its effective radius, defined as the radius
of the isophote that encloses half of the total stellar luminosity in projection.
In our simulations, we adopt the effective radius $R_e$ as the projected radius enclosing half of the stellar mass.
Here we assume a constant stellar mass to light ratio.
From Eqn. (\ref{eq:hern_star}), the effective radius is related to the quarter- and half-mass radii by $R_e = 1.8153r_{1/4}=0.752r_{1/2}$
(see also \citealt{1998gaas.book.....B}).
We remove particles of the two-component Hernquist profile outside $r \simeq 20 a_{dm}$ in our numerical realizations in order to avoid the occurrence of a very small number of very distant particles.
This choice of outer radius does not affect the half-mass radius significantly.
If the total stellar mass and the scale radius are specified, the Hernquist profile determines a unique structure.

We construct the bulge and the dark matter halo to be in equilibrium in the total gravitational potential by an N-body method. We assume that both components are both spherical and isotropic.
The particle positions for each component are initialized from the density profile, while the particle velocities are drawn based on each component's equilibrium distribution function $f_i(E)$ (\citealt{2008gady.book.....B}),

\begin{equation}
  \label{eq:eddington}
  f_i (E) = \frac {1} {\sqrt{8} \pi^2} \int_0^{E} \frac {\mathrm{d}^2 \rho_i} {\mathrm{d}\psi^2} \frac {\mathrm{d}\psi} {\sqrt{E-\psi}},
\end{equation}
where $E$ is the relative energy of each particle of component $i$,
 $\rho_i$ is the density profile of component $i$, 
 and $\psi$ is the total gravitational potential.
\citet{1996ApJ...471...68C} derives the analytical distribution function of the two-component Hernquist profiles.
We use this formula to obtain the particle velocity distribution.

Here, we only briefly summarize our N-body realization method, which is described in detail in \citet{1994MNRAS.269...13K}.
A particle's position is first determined by sampling from the density distribution.
Next the particle's velocity vector is randomly sampled from inside a velocity-space sphere whose radius is the local escape velocity, $v_{esc}=\sqrt{-2\psi}$.
Finally, based on the distribution function of the bulge or halo component, we determine whether we select the velocity vector or not using the acceptance-rejection technique.

\begin{table*}
  \centering
  \begin{minipage}{145mm}
  \renewcommand{\thempfootnote}{\fnsymbol{mpfootnote}}
    \caption{Initial galaxy models for dry merger simulations. 
    First column: name of the initial galaxy model,
    Model A is the compact massive ETG seen at high-z. This is the primary galaxy in our minor merger simulations. Model B is the compact, less massive ETG. This is used as a compact satellite. Model C is the less massive ETG that obeys the stellar mass-size relation in the local Universe. Model D is the most compact satellite (see \S\ref{subsec:model}).
    From column 2 to column 10: dark matter halo mass; redshift; halo concentration; virial radius of the halo measured at each redshift; number of dark matter particles; stellar mass; effective radius; number of stellar particles; softening length.
    }
    \label{table:initial}
    \begin{tabular}{@{}cccccccccc@{}}
      \hline
      \hline
      Model & $M_{dm}$($M_{\odot}$) & z & c & $r_{vir}$(kpc) & $N_{dm}$ &$M_{*}$($M_{\odot}$) & $R_{e}$(kpc) & $N_{*}$ & $\epsilon$(kpc) \\
      \hline
      A  & $10^{12}$ & 2.0 & 10 & 105 & $6.0\times10^{5}$ & $10^{11}$ & 1.0 & $6.0\times10^{4}$ & 0.03 \\
      B  & $10^{11}$ & 2.0 & 10 & 48.8 & $6.0\times10^{4}$ & $10^{10}$ & 0.464 & $6.0\times10^{3}$ & 0.03 \\
      C  & $10^{11}$ & 0.0 & 10 & 120 & $6.0\times10^{4}$ & $10^{10}$ & 1.15 & $6.0\times10^{3}$ & 0.03 \\
      D  & $10^{11}$ & 2.0 & 10 & 48.8 & $6.0\times10^{4}$ & $10^{10}$ & 0.148 & $6.0\times10^{3}$ & 0.03 \\
      \hline
    \end{tabular}
  \end{minipage}
\end{table*}

\begin{table*}
  \centering
  \begin{minipage}[t]{180mm}
    \caption{Physical properties of the merger remnants within the fixed radius, 60kpc, in each run.
    The properties of the initial galaxy are defined within 60kpc.
    First column: name of the simulation, 2A: major merger of model A, 1A: the primary galaxy, 1/5/10: the number of minor mergers, B/C/D: type of the satellite galaxy, sq/sm: sequential or simultaneous minor mergers.
    From column 2 to column 10: stellar mass; virial radius of the dark matter halo; half-mass radius of the stellar system; effective radius; average density within 1kpc; average density within the effective radius; velocity dispersion; size growth efficiency; efficiency of the change of velocity dispersion (see \S\ref{sec:r_e_sigma_e}).    
     }
    \label{table:simulation60_new}
    \begin{tabular}{@{}ccccccccccc@{}}
      \hline
      \hline
      Name & stellar mass & $r_{\mathrm{vir}}$ & $r_{*,\mathrm{half}}$ & $R_e$ & $\bar{\rho} (<1\mathrm{kpc}) $ & $\bar{\rho} (<R_e)$ & $\sigma_{e}$  &  $\alpha$  &  $\gamma$  \\
       & ($M_{\odot}$) & (kpc) & (kpc) & (kpc) & $ (10^{10} M_{\odot} \mathrm{kpc}^{-3}) $ & $(10^{10} M_{\odot} \mathrm{kpc}^{-3})$ & (km/s)  &  $(R_e \propto M_{*}^{\alpha})$  &  $(\sigma \propto M_{*}^{-\gamma})$  \\

      \hline
      A (initial)    & $9.82\times10^{10}$  & 257 & 1.29 & 0.968   & 0.992 & 1.07     & 262   &  \\
      \hline
      2A               & $1.91\times10^{11}$ & 309 & 3.81 & 2.81 & 0.933 & 0.0864  & 253   &  1.60  &  0.0528  \\
      \\
      1A1B   & $1.04\times10^{11}$ & 261 & 1.50 & 1.12 & 0.961 & 0.737 & 256  &  2.60  &  0.435\\
      1A1C   & $1.02\times10^{11}$ & 260 & 1.40 & 1.05 & 0.981 & 0.878 & 258  &  2.37  &  0.468\\
      1A1D   & $1.05\times10^{11}$ & 262 & 1.55 & 1.16 & 0.935 & 0.667 & 253  &  2.69  &  0.501\\
      \\
      1A5Bsq   & $1.32\times10^{11}$ & 275 & 2.90 & 2.15 & 0.852 & 0.137 & 237  &  2.72  &  0.350\\
      1A5Csq   & $1.16\times10^{11}$ & 268 & 1.92 & 1.44 & 0.951 & 0.400 & 249  &  2.44  &  0.310\\
      1A5Dsq   & $1.35\times10^{11}$ & 276 & 2.77 & 2.08 & 0.696 & 0.149 & 223  &  2.38  &  0.497\\
      \\
      1A10Bsq   & $1.49\times10^{11}$ & 301 & 4.13 & 3.00 & 0.853 & 0.0581 & 226  &  2.71  &  0.351  \\
      1A10Csq   & $1.19\times10^{11}$ & 286 & 2.04 & 1.54 & 0.951 & 0.339 & 247  &  2.42  &  0.318\\
      1A10Dsq   & $1.70\times10^{11}$ & 302 & 5.14 & 3.76 & 0.677 & 0.0339 & 206  &  2.46  &  0.440\\
      \\
      1A10Bsm  & $1.60\times10^{11}$ & 294 & 3.78 & 2.78 & 0.873 & 0.0767   & 241  &  2.17  &  0.175  \\
      1A10Csm  & $1.54\times10^{11}$ & 287 & 3.88 & 2.86 & 0.964 & 0.0706   & 243  &  2.42  &  0.166\\
      \hline
      \hline
    \end{tabular}
  \end{minipage}
\end{table*}

\begin{table*}
  \centering
  \begin{minipage}[t]{130mm}
    \caption{Physical properties of the merger remnants within $r_{\mathrm{trunc}} = 30$, 60, and 90kpc in binary minor merger simulations.
    Each column is similar to Table \ref{table:simulation60_new}.
    }
    \label{table:define_remnant}
    \begin{tabular}{@{}ccccccc@{}}
      \hline
      \hline
      Name & stellar mass &  $R_e$ & $\sigma_{e}$ & $\alpha$  &  $\gamma$  \\
       & ($M_{\odot}$) & (kpc) & (km/s)  &  $(R_e \propto M_{*}^{\alpha})$  &  $(\sigma \propto M_{*}^{-\gamma})$  \\

      \hline
      $r_{\mathrm{ana}} = 30$kpc  \\
      \hline
      1A1B   & $1.01\times10^{11}$ & 1.06 & 257  &  2.71  &  0.535\\
      1A1C   & $0.984\times10^{11}$ & 0.990 & 259  &  2.63  &  0.758\\
      1A1D   & $1.02\times10^{11}$ & 1.11 & 254  &  2.79  &  0.569\\
      \hline
      $r_{\mathrm{ana}} = 60$kpc  \\
      \hline
      1A1B   & $1.04\times10^{11}$ & 1.12 & 256  &  2.60  &  0.435\\
      1A1C   & $1.02\times10^{11}$ & 1.05 & 258  &  2.37  &  0.468\\
      1A1D   & $1.05\times10^{11}$ & 1.16 & 253  &  2.69  &  0.501\\
      \hline
      $r_{\mathrm{ana}} = 90$kpc  \\
      \hline
      1A1B   & $1.05\times10^{11}$ & 1.14 & 256  &  2.55  &  0.408\\
      1A1C   & $1.03\times10^{11}$ & 1.07 & 257  &  2.28  &  0.457\\
      1A1D   & $1.06\times10^{11}$ & 1.18 & 252  &  2.67  &  0.534\\
      \hline
    \end{tabular}
  \end{minipage}
\end{table*}

\subsection{Model parameters}\label{subsec:model}

For the progenitor galaxies in major and minor merger simulations, we assume four galaxy models: a massive ETG, two less massive compact ETGs, and a less massive diffuse ETG.
Parameters of these galaxy models are summarized in Table \ref{table:initial}.
We assume that the massive ETG (model A) has the stellar mass $M_{*}=10^{11}M_{\odot}$, with an effective radius $R_{e}=1.0$kpc, which corresponds to the scale radius $a=0.551$kpc.
These properties are consistent with that from observations of the high-z compact massive ETGs (e.g. \citealt{2007MNRAS.382..109T}).
They are embedded in dark matter haloes of $M_{dm}=10^{12}M_{\odot}$ and $r_{vir}$ at $z=2.0$.
In this paper, we assume dark matter to stellar mass ratio $M_{dm}/M_{*}=10$ in all models.
We also assume that for all initial dark matter haloes, $c_{vir}=10$.
Many cosmological simulations  predict that the halo concentration depends on the mass and the formation redshift (\citealt{2001MNRAS.321..559B}), but we select a constant value in favour of simplicity in the model.

We assume three different satellite galaxies: compact (model B), diffuse satellites (model C), and very compact satellites (model D).
We assume that the formation epoch of model B is the same as that of model A, both occurring at $z=2.0$, and that these models are expected to have the same average density.
Based on this assumption, we construct the stellar system for model B by simply scaling down that of model A, given the mass fraction (1/10 of model A) and scale radius ($1/\sqrt[3]{10}$ of model A).
We also assume the dark matter halo of model B is $M_{dm}=10^{11}M_{\odot}$.
For model C galaxies, we assume that they have the properties of the local ETGs.
For their effective radii, we use the observed local stellar mass-size relation of early-type galaxies in SDSS (\citealt{2003MNRAS.343..978S}),
\begin{equation}
  R_{e} = 4.16 \left( \frac {M_{*}} {10^{11} M_{\odot}} \right)^{0.56} \mathrm{kpc}.
\end{equation}
The model C galaxies have dark matter haloes with $M_{dm}=10^{11}M_{\odot}$ and $r_{vir}$ at $z=2.0$. 
Recent observations indicate that early-type galaxies follow a stellar mass-size relation $\log (R_{e}) = S + T\log(M_{*})$ with $T \sim 0.6 - 0.7$, not significantly dependent on redshift, and $S$ increasing for decreasing redshift (e.g. \citealt{2012ApJ...746..162N}).
From this, we assume that model A and model D lie on the same stellar mass-size relation with $T = 0.57$, as in \citet{2012ApJ...746..162N}, giving model D satellites that are more compact than both model B and C.
We examine effects of the stellar densities of the satellites on the properties of merger remnants using this set of satellite models.

\subsection{Simulations and test run}\label{subsec:testrun}
We use the GADGET-2 code (\citealt{2005MNRAS.364.1105S}), an efficient parallelized treecode, to perform our simulations.
We set the tolerance parameter $\alpha=0.001$ for the relative cell-opening criterion in the tree algorithm, the timestep tolerance parameter $\eta=0.005$, and the force softening $\epsilon=0.03$ kpc for both star and dark matter particles.
With this choice, the minimum timestep is $< 0.02 \mathrm{Myr}$.
In this study, we use $6.0\times10^5$ dark matter particles and $6.0\times10^4$ stellar particles for model A.
For model B and model C, we use $6.0\times10^4$ dark matter particles and $6.0\times10^3$ stellar particles.
Each halo and stellar particle has the same mass.

\begin{figure}
  \resizebox{84mm}{!}{\input{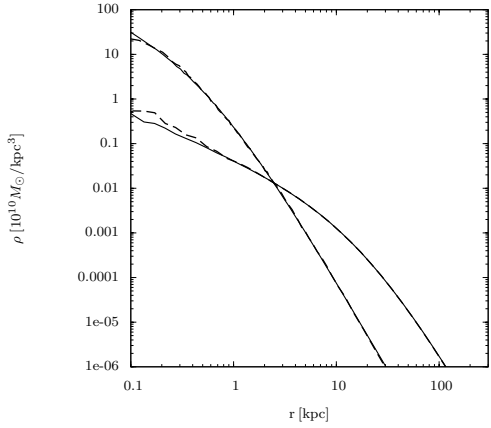}}  
  \caption{Density profiles of the stellar system and the dark matter halo of a single model A galaxy (Table \ref{table:initial}) simulated as a test run to check whether our two-component galaxy model is stable.
  The solid line shows the initial profile ($t=0$).
  The dashed line shows the profile after 3.6 Gyr.
  The concentrated ones represent the stellar system.
  }
  \label{fig:test3}
\end{figure}

We should draw attention to the effect of two-body relaxation, since the dynamical time of the compact massive ETGs is relatively short compared with the galaxy merging time-scale, and we need the simulation time to be much larger than the dynamical time of the stellar system.
We have confirmed that our two-component galaxy model is stable at all radii $r > a = 0.551 \mathrm{kpc}$ for $\sim 3.6$ Gyr of the simulation.
We define the half-mass dynamical time of a stellar system as
\begin{equation}
  t_{dyn} = \sqrt{ \frac {3\pi}{16G\rho_{1/2}}} = \pi \sqrt{ \frac {r_{1/2}^3}{2GM_{*}}},
\end{equation}
where $\rho_{1/2}=3M_{*}/8\pi r_{1/2}^3$ is the mean density inside the half-mass radius $r_{1/2}$.
Here, we consider only the stellar component for simplicity.
We estimate the two-body relaxation time-scale $t_r$ from the following expression,
\begin{equation}
  t_r(r) \simeq \frac{0.1N}{\ln N} t_{dyn},
\end{equation}
where $N$ is the number of stellar particles in a radius $r$.
The half-mass dynamical time of model A is $t_{dyn} \simeq 5.2 $ Myr and the relaxation time is $t_r \simeq 2$ Gyr which is comparable to the total computational time of each of our simulation sets.
To check whether our two-component galaxy model is stable we simulated a single model A galaxy for $\sim 3.6$ Gyr as a test run.
Fig. \ref{fig:test3} shows several snapshots of the density profiles of the stellar system.
This figure clearly shows that our model is stable in all radii $r \ga 0.2 \mathrm{kpc}$.
We checked the conservation of the total energy $E$ and the virial ratio ($V=2T/W$), obtaining $|\Delta E/E| \la 1$ percent and $|\Delta V/V| \la0.35$ percent. 
We also checked the stellar half-mass radius, the stellar effective radius and the stellar velocity dispersion at 3.6Gyr, and we 
confirmed that changes of both the half-mass radius and the effective radius are less than 3 percent, and the projected velocity dispersion is nearly constant.
These changes are smaller than that due to mergers in all runs (see Table \ref{table:simulation60_new}).
In order to avoid the influence of numerical error, we stopped our simulations at $3.6$ Gyr.

\subsection{Production run set}

We have run N-body simulations of dry major and minor mergers to investigate the effects of mergers on the evolution of size and velocity dispersion in ETGs.
In this paper, we mainly focus on the dry minor merger cases.
We also simulate a dry major merger for comparison.
We describe the setup of our simulations below.
Table \ref{table:simulation60_new} summarizes the parameters used in our run set.

Merger orbits are critical to the properties of the merger remnants.
We assume head-on parabolic mergers with zero orbital angular momentum in all our simulations.
In \citet*{2006MNRAS.369.1081B}, who studied dry major mergers of early-type galaxies, merger remnant properties are highly influenced by orbital parameters of the mergers.
They showed that the stellar bulges of the merger remnants in the cases of radial orbits are larger than in the cases of circular orbits.
Therefore, our simulations are optimal cases for the size growth of galaxies.

\subsubsection{Major mergers}

In run 2A, we simulate a major merger for the parabolic and head-on encounter of same initial galaxies of model A.
The initial separation between the two galaxies is 200kpc, and the initial relative speed is 308km/s.
The merger appears to be mostly completed by 2.3 Gyr.

\subsubsection{Minor mergers}

We simulate two cases of minor mergers, involving simultaneous minor mergers and sequential ones.
We investigate how the merger remnants are affected by a different types of minor mergers.
In each case, we assume three different model galaxies (model B, model C, and model D, as in Table \ref{table:initial}).
Initially, we assume the central galaxy of model A (the ``primary galaxy'') is surrounded by satellite galaxies of model type B, C or D.
In the simultaneous minor mergers case, the satellites are randomly distributed around the primary galaxy with the relative distance between each satellite and the primary galaxy in the range [130, 170] kpc.
During the merging events, the satellites are accreted at the same time.
The merger orbit of each satellite is assumed to be parabolic and head-on in the absence of the other satellites.
We calculate the relative velocity using the total mass of the progenitors before removing very distant particles.

As shown in \S 2, in hierarchical structure formation scenario in the $\Lambda$CDM cosmology, there are many sequential minor mergers. We study the effects of this sequential minor merging in our next case. 
Here, each satellite merges with the primary galaxy from random directions every 0.2 Gyr.
We simulate five sequential (run 1A5Bsq, 1A5Csq, 1A5Dsq) and then ten sequential (run 1A10Bsq, 1A10Csq, 1A10Dsq) minor mergers.
In these cases, the period of the sequential minor mergers roughly corresponds to the time interval from $z=2$ to $z=1$.
We assume the three types of satellites (model B, model C, model D), and compare the properties of the merger remnants.
In the sequential minor mergers, the mass ratio of the first merger $M_{2}/M_{1} = 1/10$, that of the second merger is 1/11, and so on.
The mass ratio can reach $\sim 1/20$ in the last merger.
Thus, the typical mass ratios of the minor mergers are $1/20 < M_{2}/M_{1} \lid 1/10$.
We also simulate single minor mergers (run 1A1B, 1A1C, 1A1D) to compare these results with the sequential minor mergers.
In these test simulations, we assume the initial relative velocity between progenitors using the total mass of the progenitors after we have removed very distant particles to ensure that the satellite orbits are strictly parabolic and head-on.

Numerical simulations were run on Cray XT4 at Center for Computational Astrophysics, CfCA,
of National Astronomical Observatory of Japan.

\subsection{Definition of merger remnants}
In this subsection, we describe our analyzing methods for the merger remnants.
In the major merger simulation (run 2A), the final state is defined when the global virial ratio of the system reaches an almost constant value.
In the case of the minor mergers, it takes more time for the minor mergers to attain complete relaxation, since the dynamical friction time-scale of the minor mergers are longer than the merging time-scale of the major merger.
Therefore, as described in \S\ref{subsec:testrun}, we stop our runs at $t = 3.6$ Gyr in the minor merger simulations, and analyze the merger remnants.
At this time, there are several surviving cores of the satellite stellar systems in the remnants from the 10 minor mergers.
However, the mass of each core is less than 5 per cent of the mass of the merger remnant which we define below.
We examine the effects of these surviving cores on properties of the merger remnants in \S\ref{sec:surf_stellar}, and find that these cores have little effect on the size growth efficiency (defined in \S\ref{sec:r_e_sigma_e}) of the remnants.

We analyse the remnant particles at the end of the simulations.
For the dark matter particles, we define the merger remnants as the particles which are gravitationally bound.
For the stellar particles, it is difficult to use that criterion, since in the minor merger simulations, the central galaxies are surrounded by very diffuse stellar particles beyond $\ga 50$ kpc that are gravitationally bound by the dark matter haloes of the remnants.
Since \citet{2010ApJ...709.1018V} use 75kpc as the robust galaxy radius to define total stellar mass in their stacked images, and determine the effective radii,
we use a radius $r_{\mathrm{trunc}} = 60$kpc to define the stellar remnants, with our reasoning described below.
We call the stellar mass within $r_{\mathrm{trunc}}$ the remnant mass.
In $r>r_{\mathrm{trunc}}$, the stellar surface density of the remnants is too low to be observed.
In fact, the surface brightness difference between 0.1kpc and 60kpc are $\Delta \mu \ga 13 \mathrm{mag \ arcsec^{-2}}$ for the remnants, assuming a constant mass-to-light ratio with radius.
Even state-of-the-art observations cannot observe such a large range of magnitudes (\citealt{2012arXiv1202.4328T}).
We examine various radii, $r_{\mathrm{trunc}}$, to find a suitable $r_{\mathrm{trunc}}$ for defining the remnants.
We have checked $R_{e}$ for $r_{\mathrm{trunc}} = 30$, 60, and 90kpc in the binary minor merger simulations as shown in Table \ref{table:define_remnant}.
We find that $R_{e}$ slightly increases from $r_{\mathrm{trunc}} = 30$kpc to $r_{\mathrm{trunc}} = 60$kpc, and is almost constant between $r_{\mathrm{trunc}} = 60$kpc and $r_{\mathrm{trunc}} = 90$kpc.
Thus, we choose $r_{\mathrm{trunc}} = 60$kpc in the following analysis of merger remnants.

We analyse the three-dimensional and projected properties of the stellar merger remnants.
The stellar half-mass radius is the radius of the sphere within which half of the mass of the remnant stellar system is enclosed.
We adopt the effective radius $R_e$ as the projected radius enclosing half of the remnant mass.
We also define line-of sight stellar velocity dispersion of the remnants as in \citet*{2005MNRAS.362..184B}.
We use the surface mass density weighted velocity dispersion within $R_e$,
\begin{equation}
  \label{vd}
  \sigma_e^2 = \frac {\int_{2\epsilon}^{R_e} \sigma_{\mathrm{los}}^2 (R) \Sigma(R)RdR} {\int_{2\epsilon}^{R_e} \Sigma(R)RdR},
\end{equation}
where $\Sigma(R)$ is the surface mass density and $\sigma_{\mathrm{los}}$ is the line-of-sight velocity dispersion.
We use 2$\epsilon$ as the lower limit of the integrations in Eqn. (\ref{vd}), where $\epsilon$ is the softening length.
Taking into account the projection effects, we average $R_e$ and $\sigma_e$ in 100 random projections.

To derive major and minor axes of the merger remnants, we obtain the principal axes of the merger remnants using the reduced moment of inertia tensor (\citealt{1991ApJ...378..496D}, \citealt{2005MNRAS.362..184B}),

\begin{eqnarray}
  I_{ij} &=& \sum_{l=1}^{N} \frac {x_{l,i}x_{l,j}}{a_l^2},\\
  a_l &=& \left(x^2 + \frac{y^2}{q^2} + \frac{z^2}{s^2} \right)^{1/2}
\end{eqnarray}
where $a_l$ is the elliptical radius and $q$ and $s$ are the axis ratios with $s\lid q\lid 1$.
$q$ and $s$ are calculated through
\begin{equation}
  q = \left( \frac{I_{yy}}{I_{xx}} \right)^{1/2} \mathrm{and} \quad s = \left( \frac{I_{zz}}{I_{xx}} \right)^{1/2},
\end{equation}
where $I_{xx}$, $I_{yy}$ and $I_{zz}$ are the principal component of the tensor.
$N$ is the number of particles in the remnants defined by all stellar particles within $r_{\mathrm{trunc}}$.
The procedure in determining these principle axes has to be done iteratively, starting with the assumption of spherical symmetry, $q=s=1$.

\section{Results}
We have run N-body simulations of the dry major, the simultaneous minor and the sequential minor mergers as described in \S\ref{sec:nbodysimulation}.
Physical quantities of the merger remnants are summarized in Table \ref{table:simulation60_new}.

\subsection{Snapshots}
\begin{figure*}
    \begin{minipage}{180mm}
        \begin{center}
            \begin{tabular}{ccc}
                \resizebox{60mm}{!}{\includegraphics{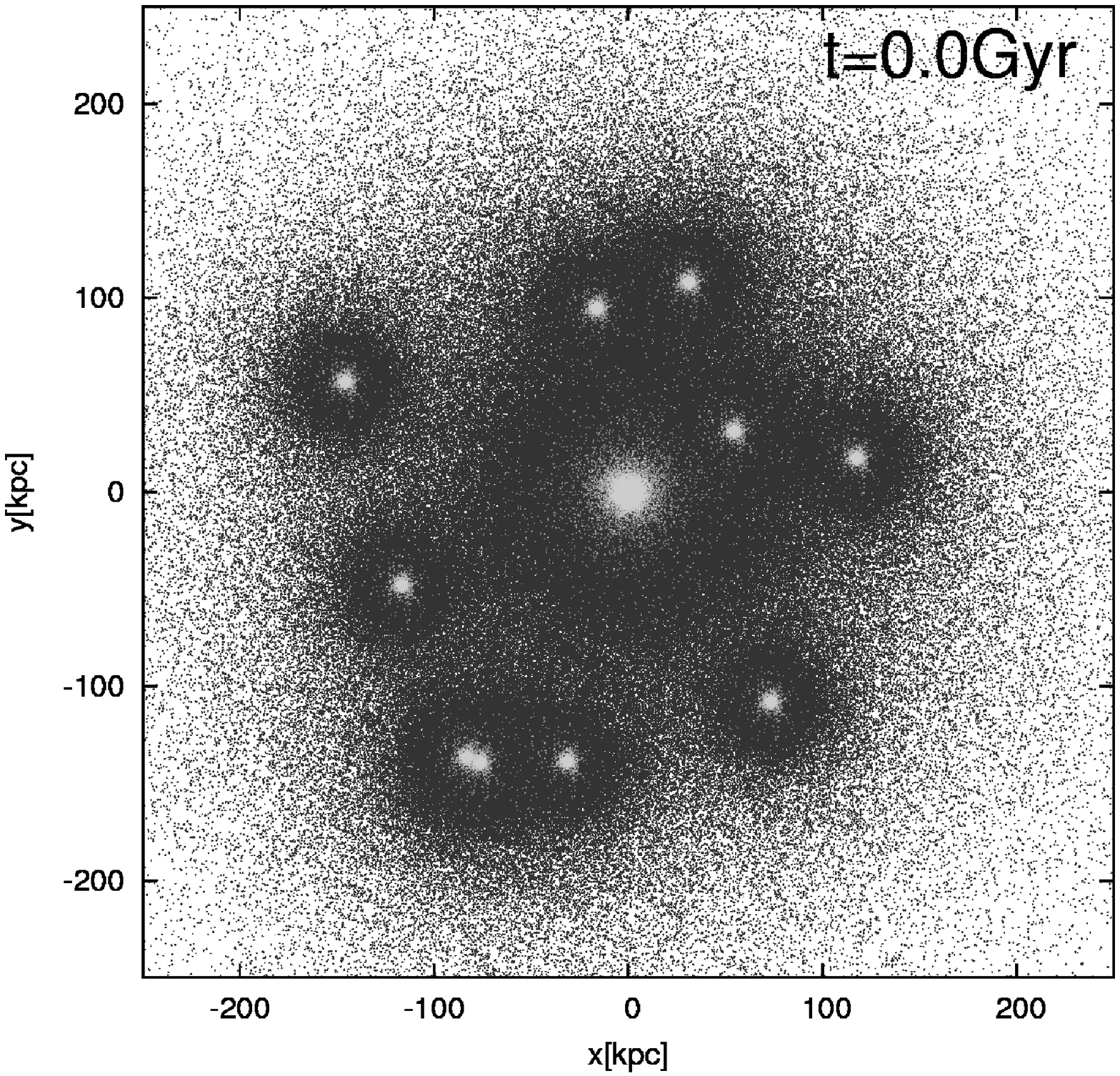}} &
                \hspace{-10mm}
                \resizebox{60mm}{!}{\includegraphics{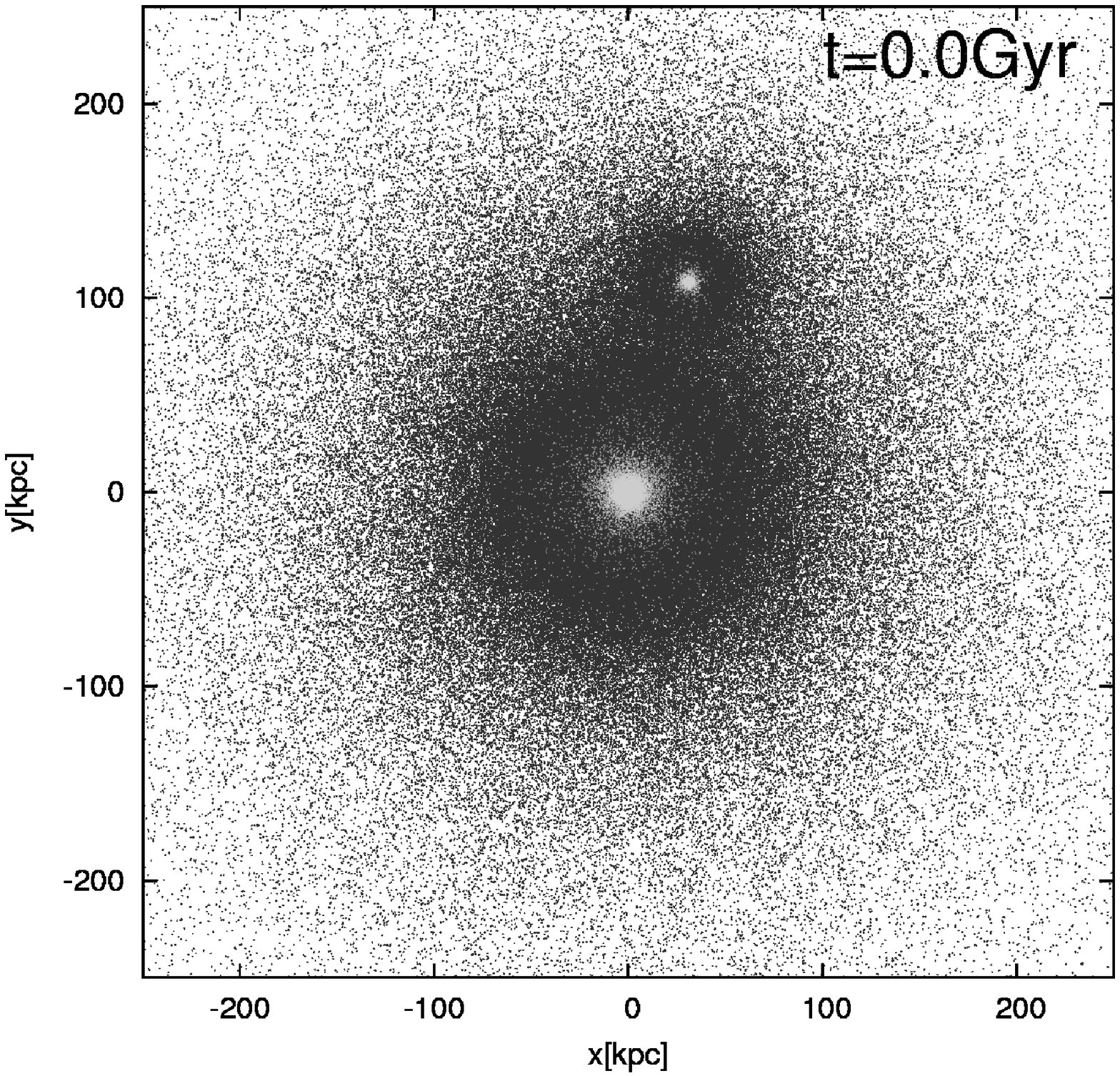}} &
                \hspace{-10mm}
                \resizebox{60mm}{!}{\includegraphics{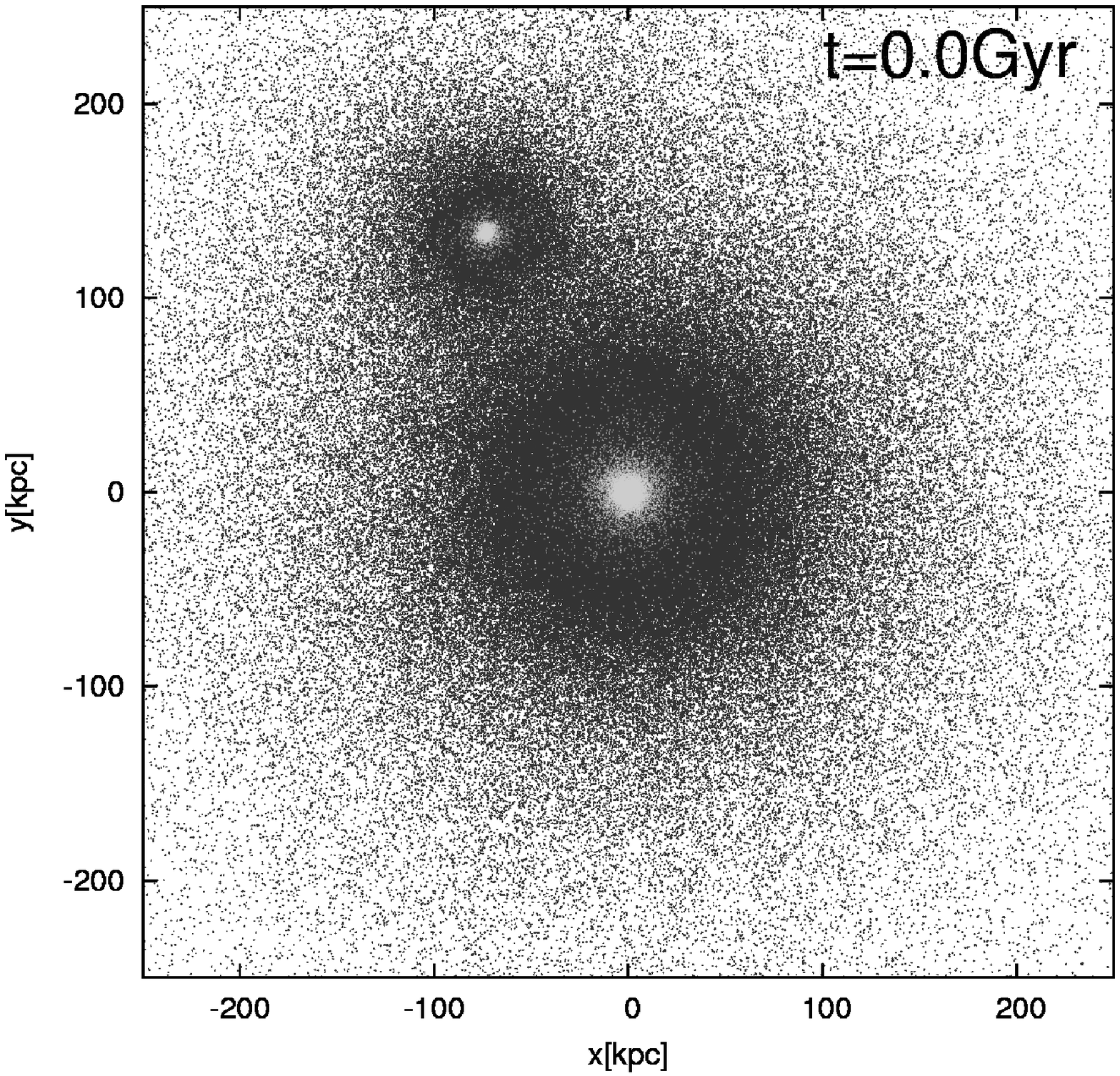}} \\

                \resizebox{60mm}{!}{\includegraphics{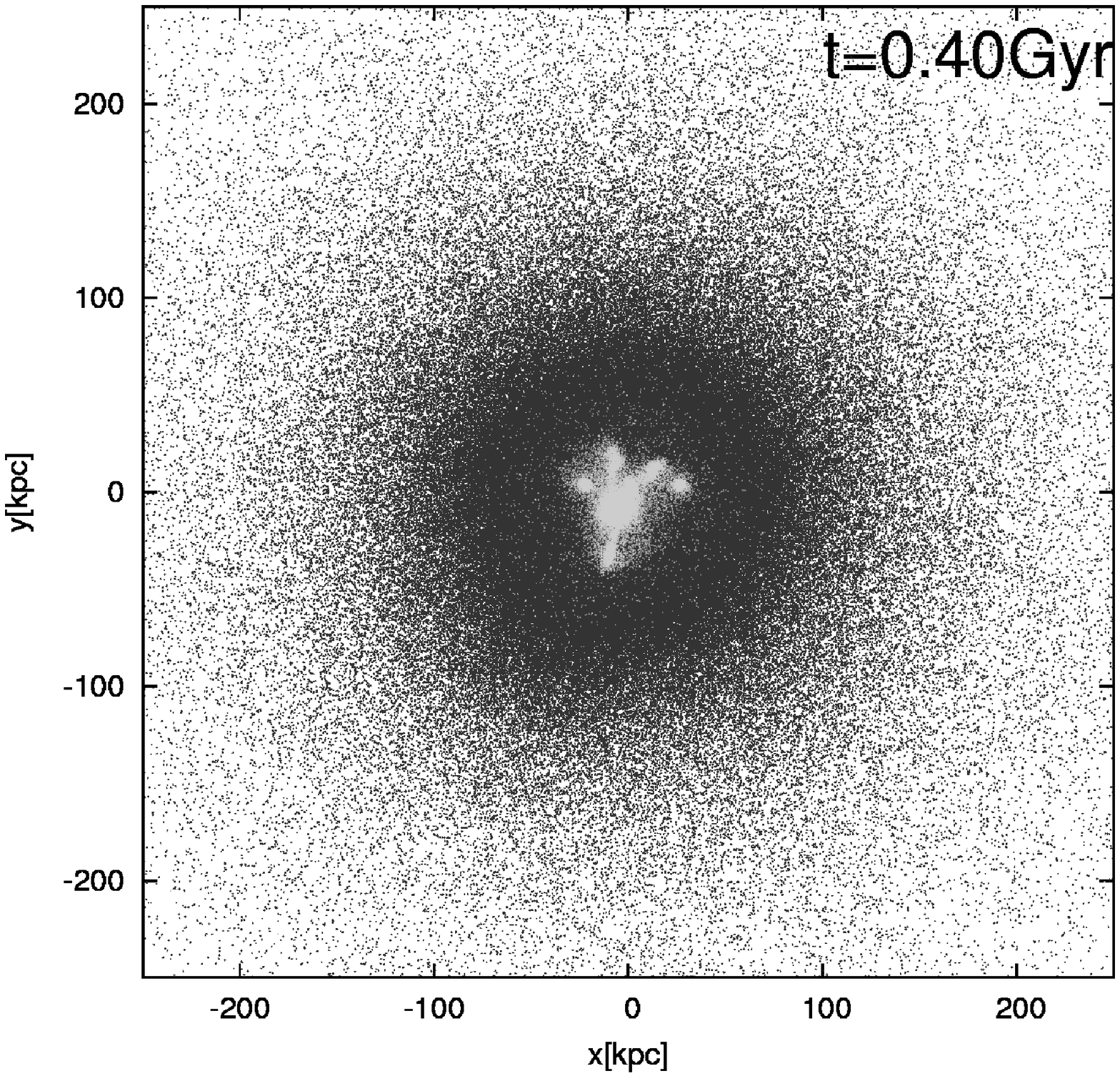}} &
                \hspace{-10mm}
                \resizebox{60mm}{!}{\includegraphics{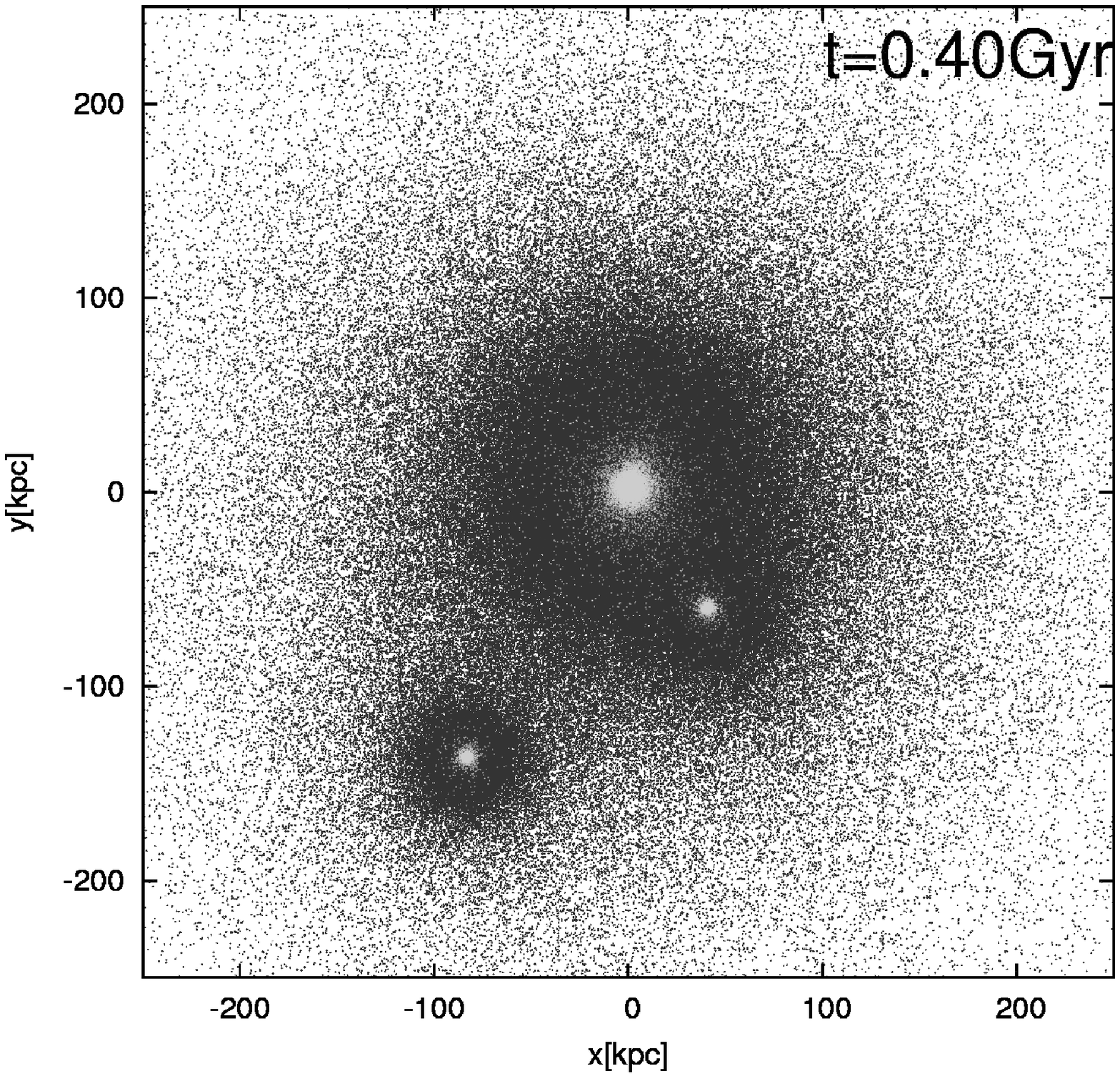}} &
                \hspace{-10mm}
                \resizebox{60mm}{!}{\includegraphics{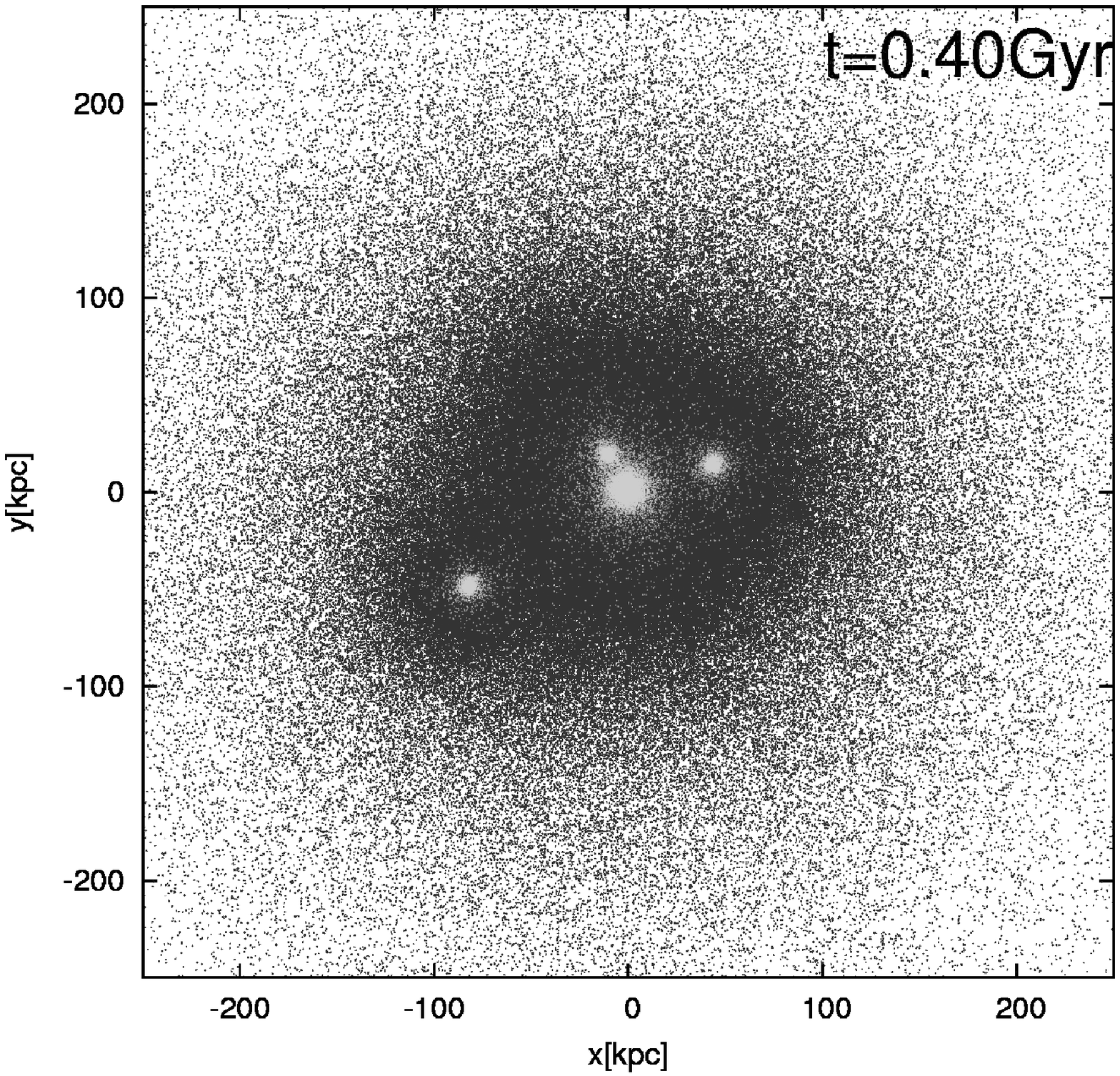}} \\

                \resizebox{60mm}{!}{\includegraphics{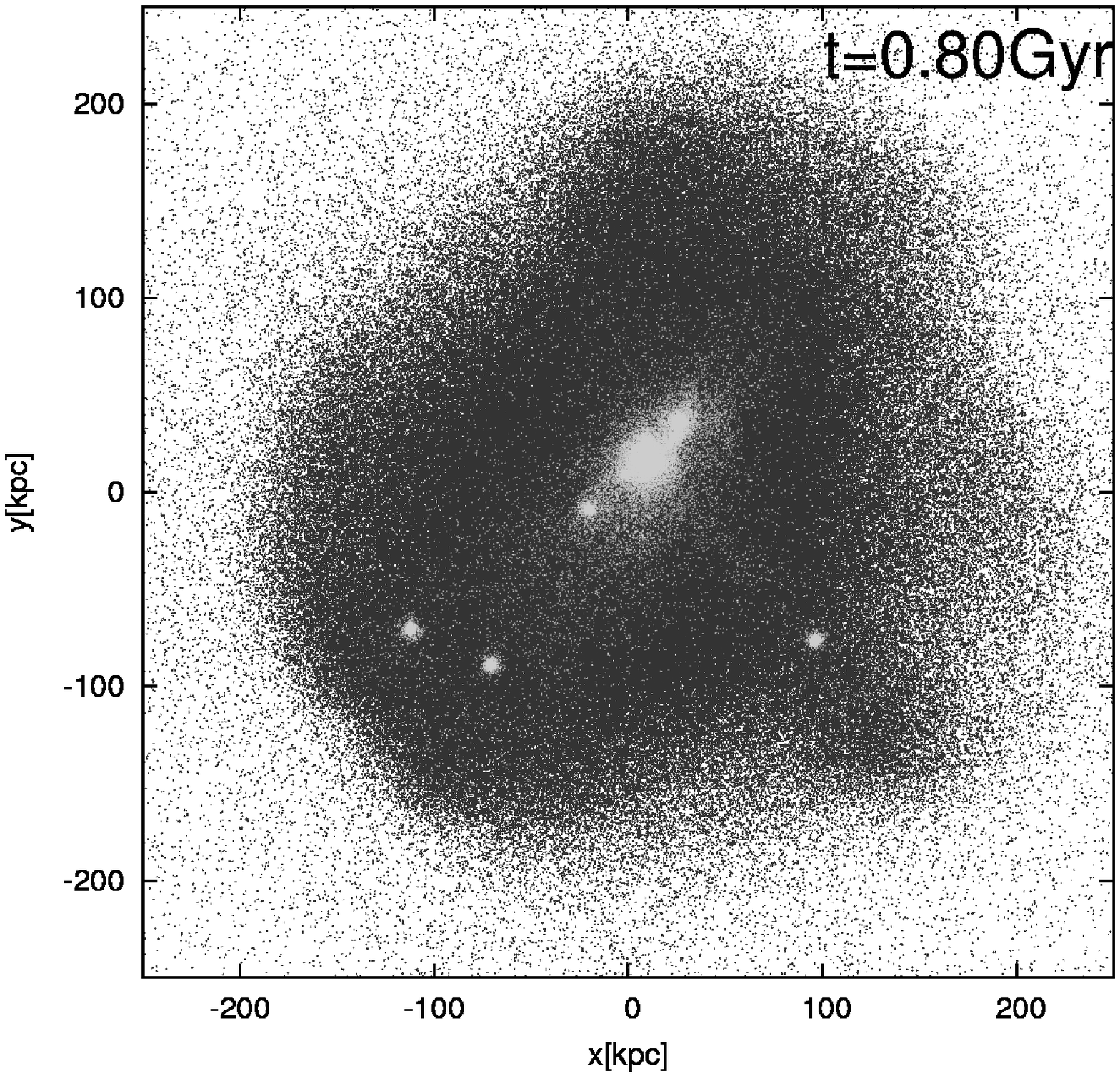}} &
                \hspace{-10mm}
                \resizebox{60mm}{!}{\includegraphics{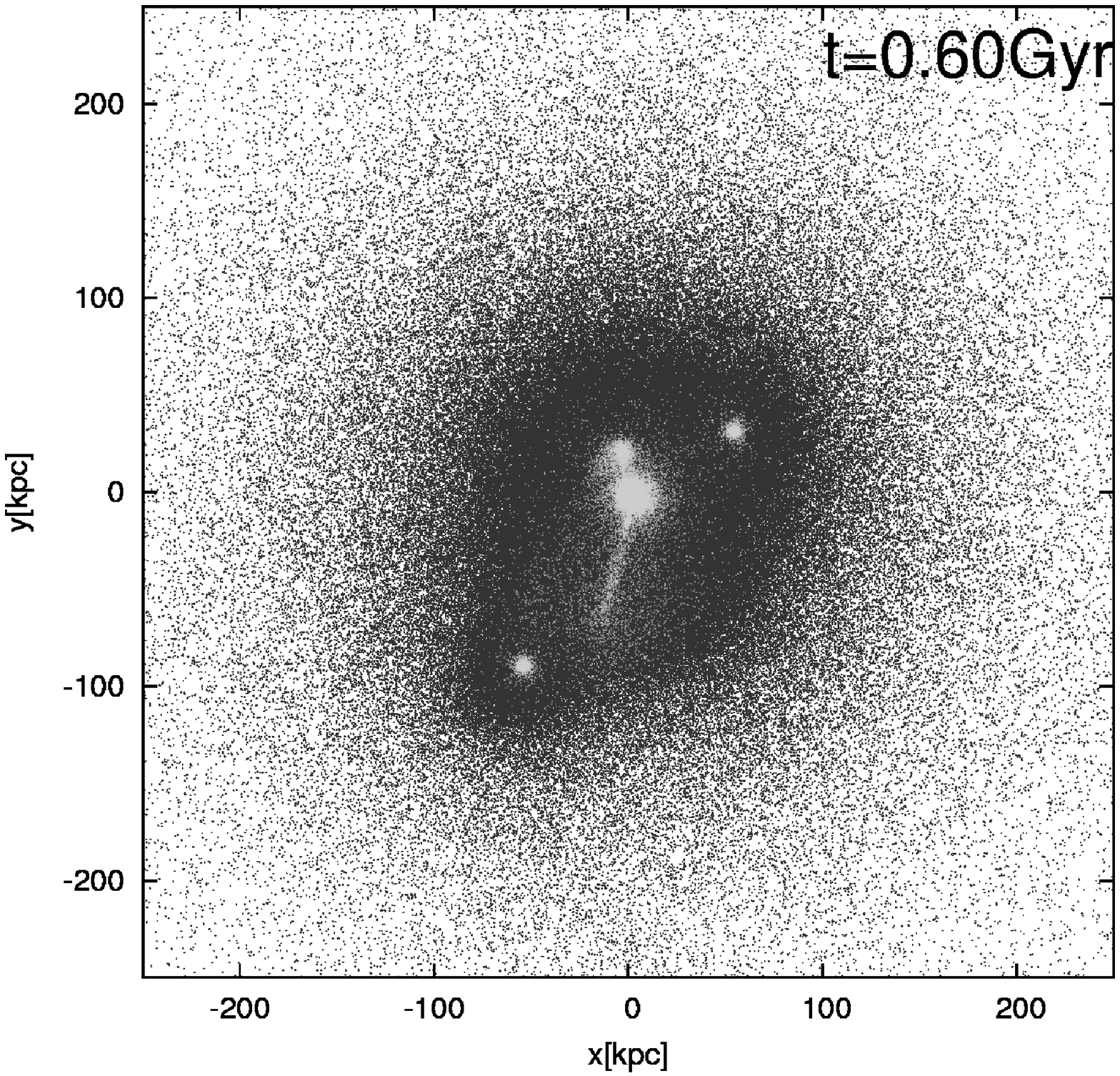}} &
                \hspace{-10mm}
                \resizebox{60mm}{!}{\includegraphics{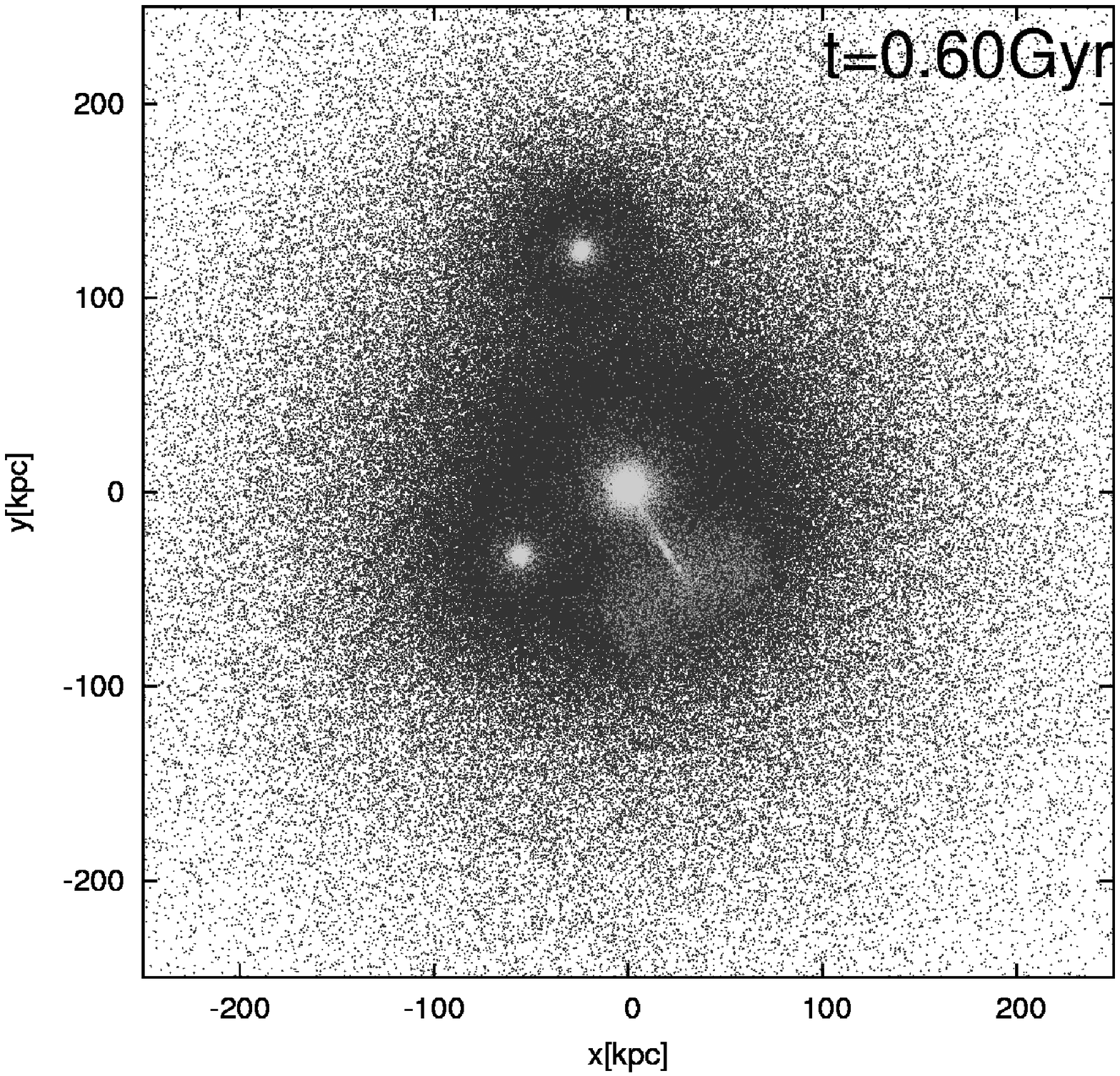}} \\

                \resizebox{60mm}{!}{\includegraphics{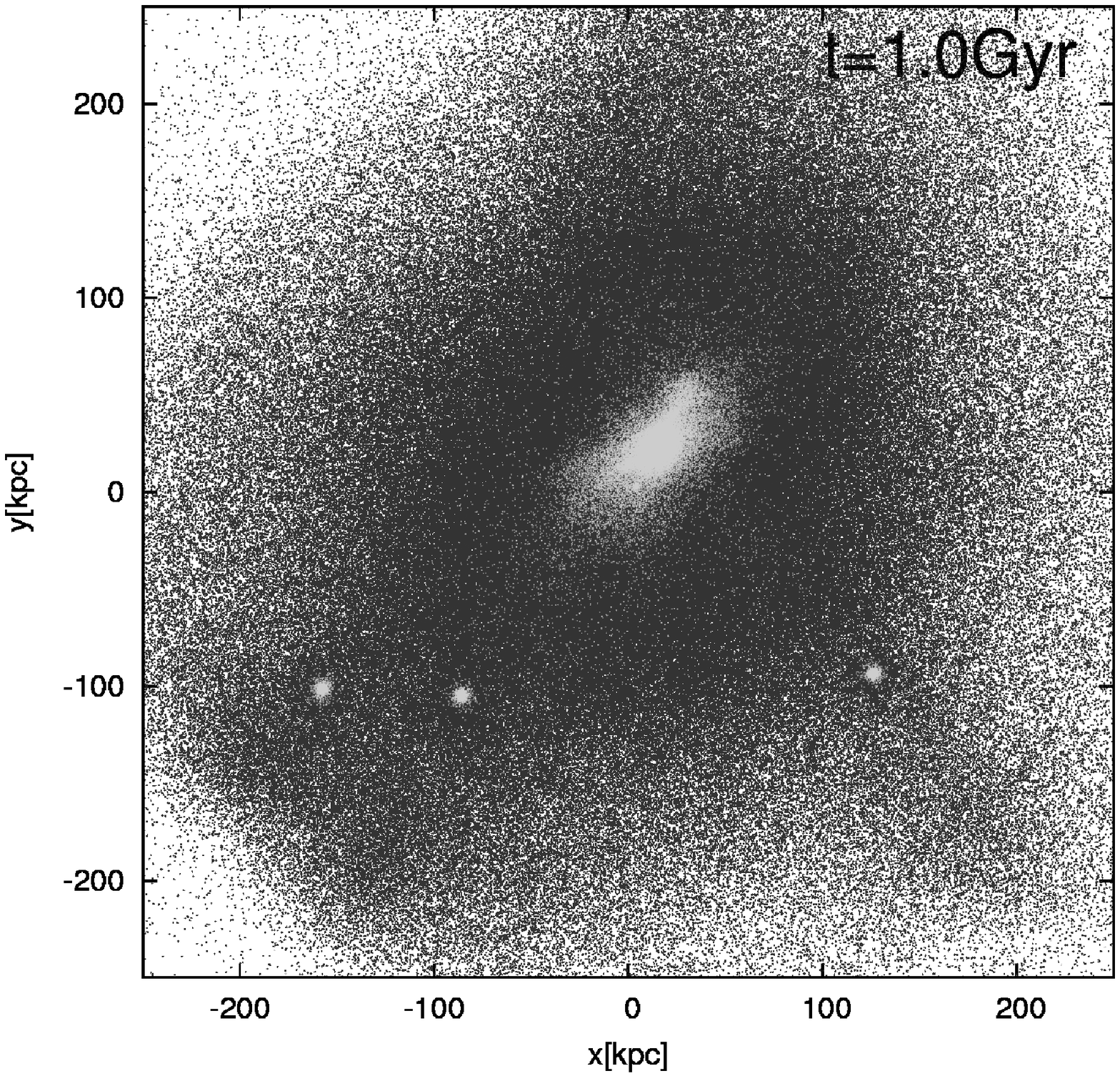}} &
                \hspace{-10mm}
                \resizebox{60mm}{!}{\includegraphics{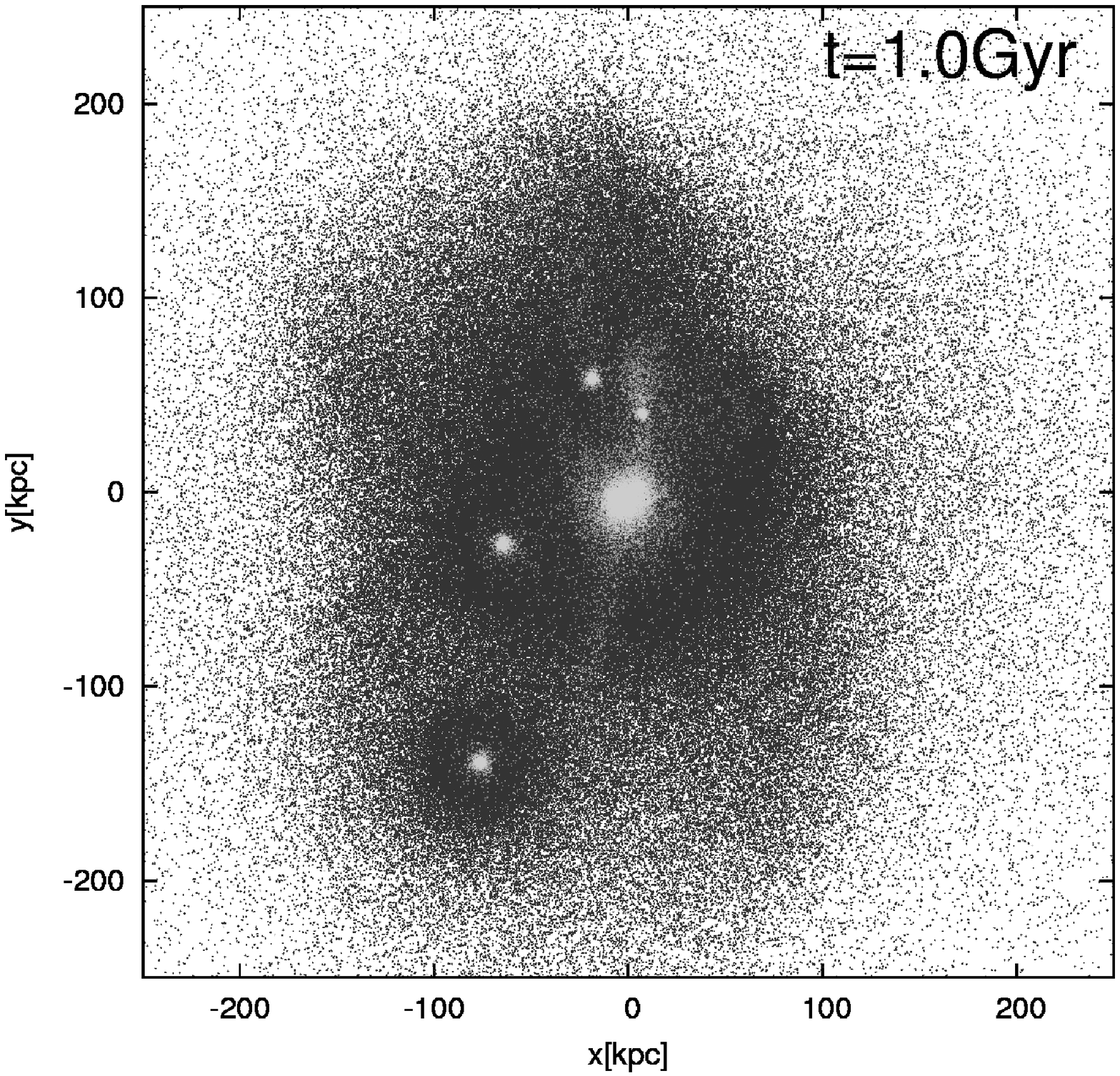}} &
                \hspace{-10mm}
                \resizebox{60mm}{!}{\includegraphics{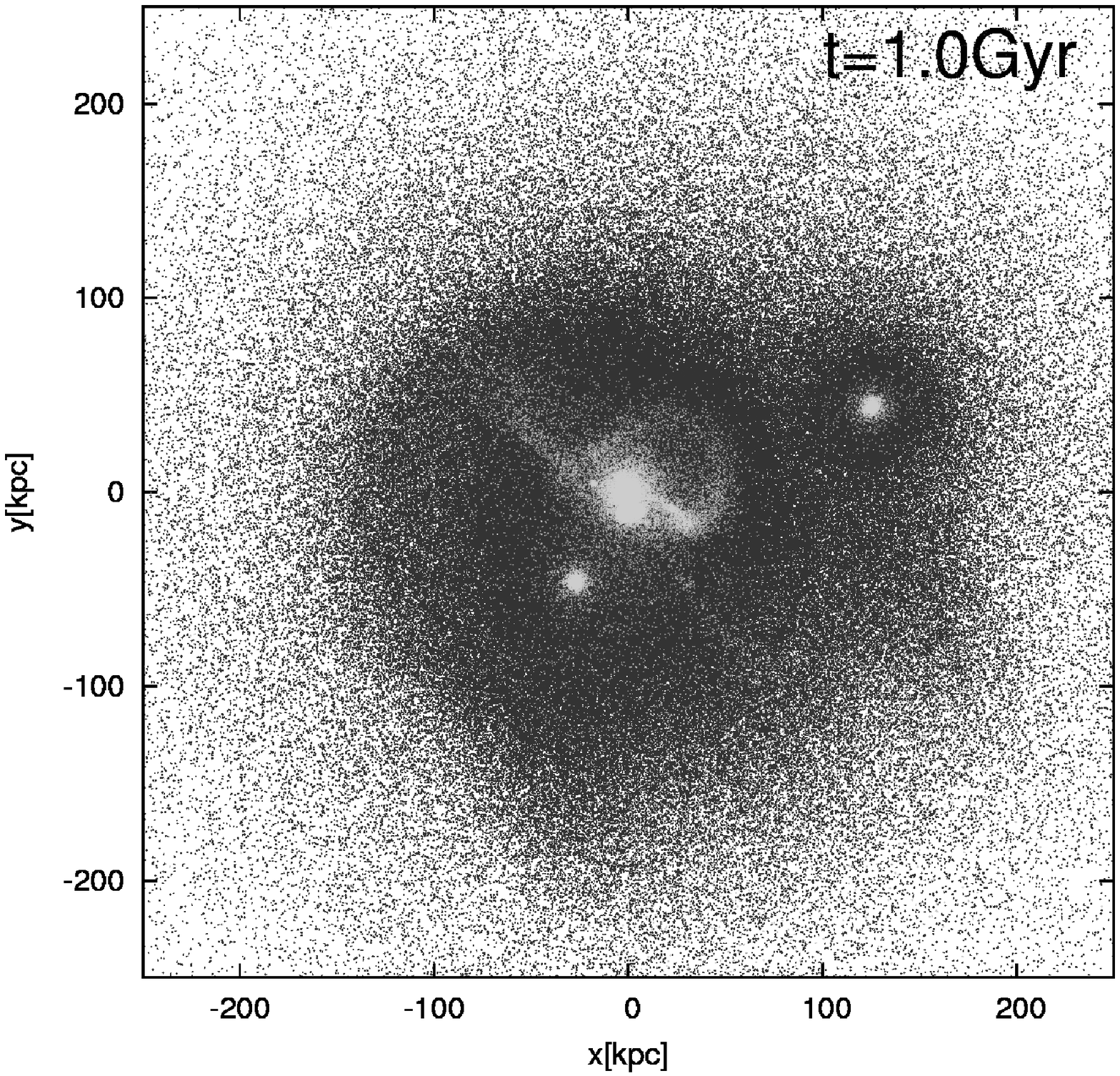}} \\
                
                \resizebox{60mm}{!}{\includegraphics{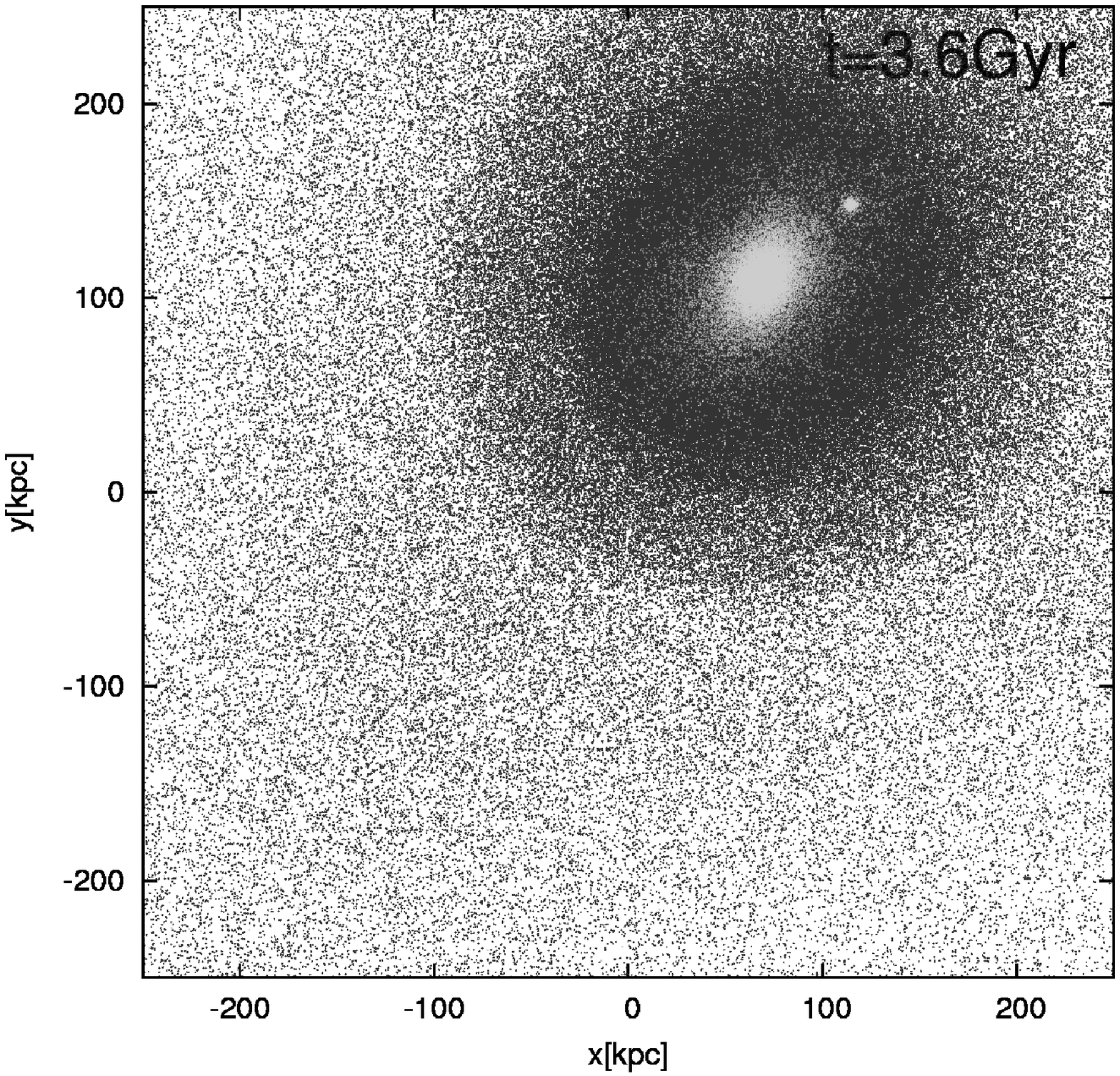}} &
                \hspace{-10mm}
                \resizebox{60mm}{!}{\includegraphics{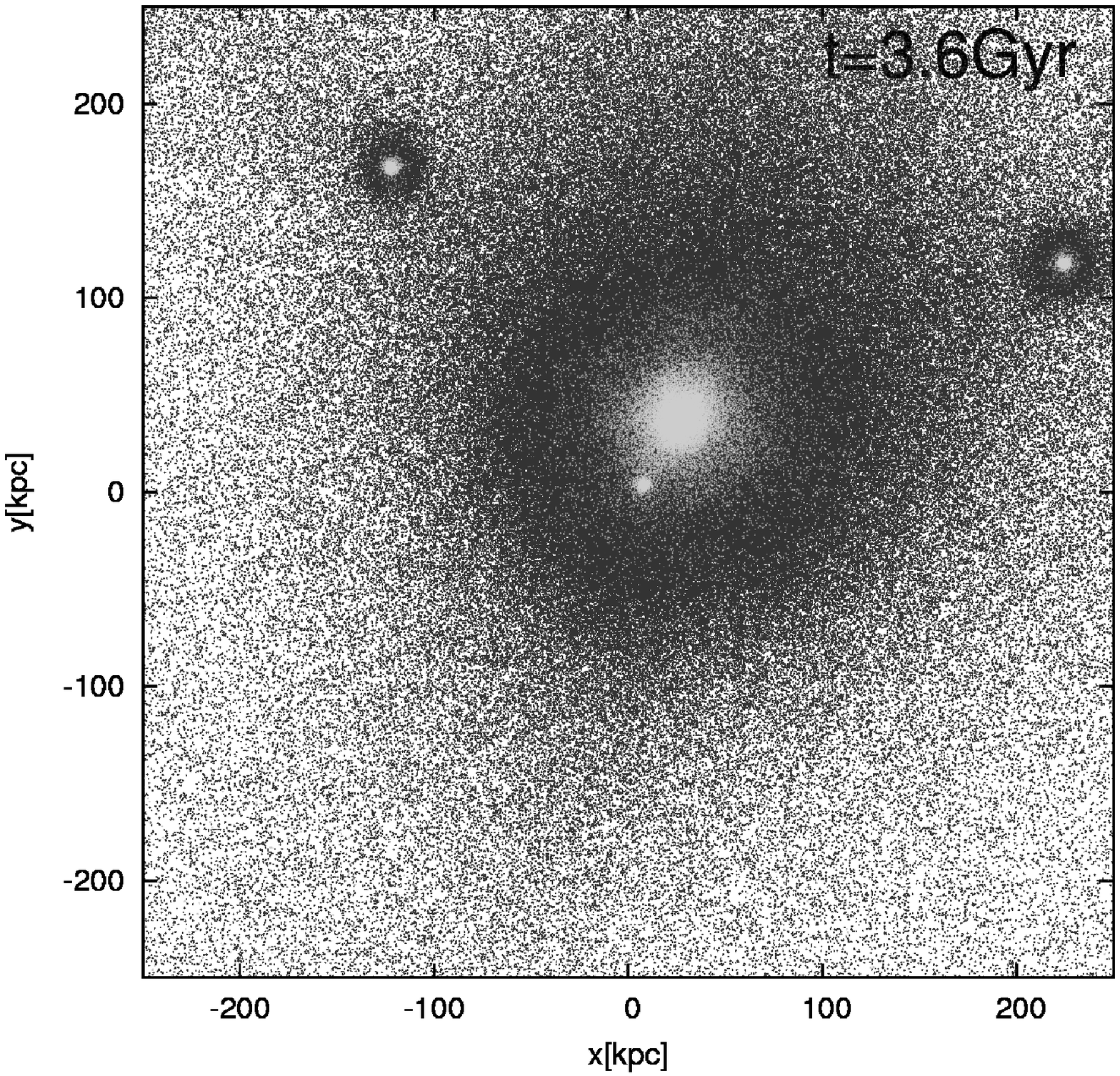}} &
                \hspace{-10mm}
                \resizebox{60mm}{!}{\includegraphics{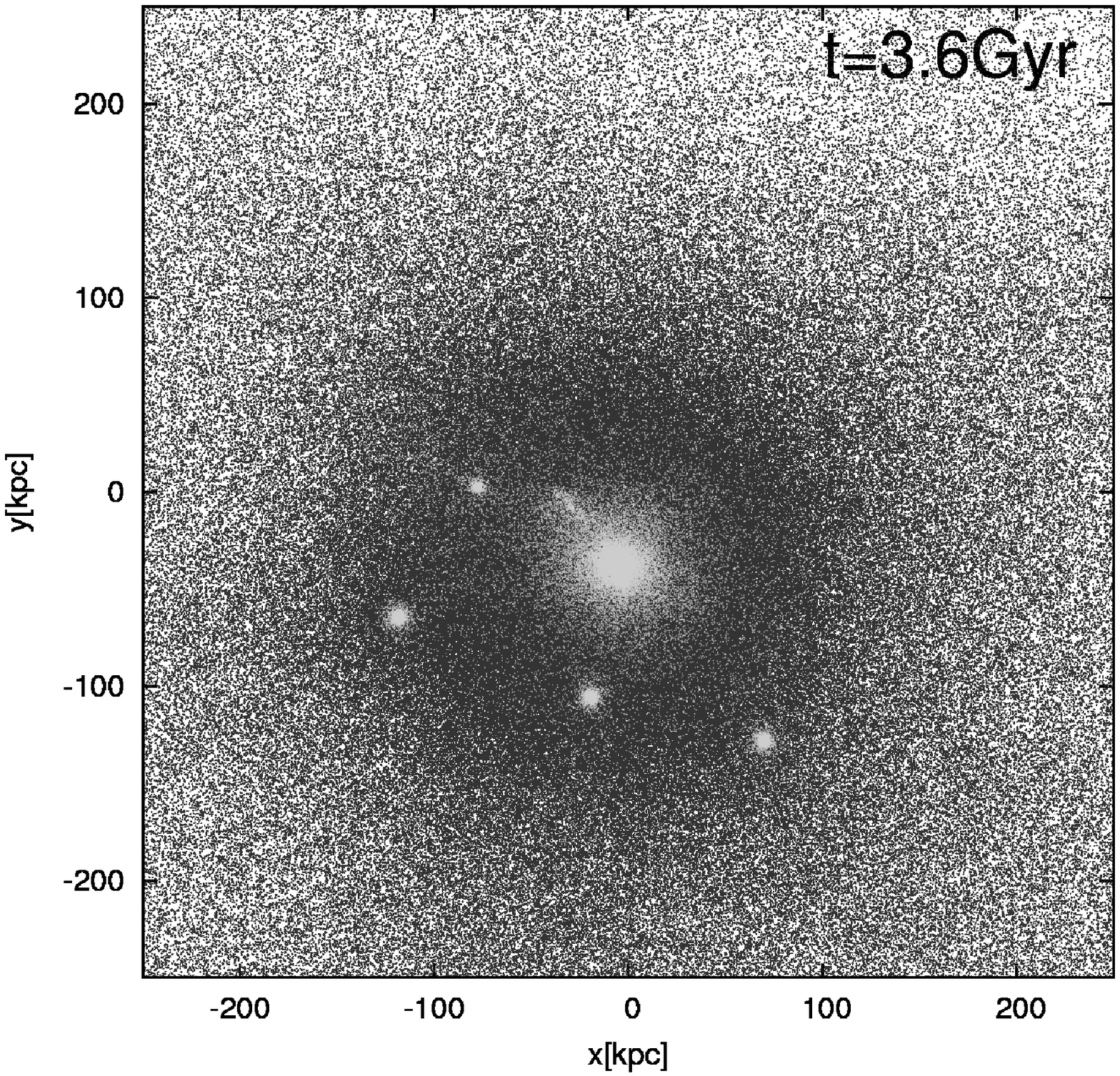}} \\
            \end{tabular}
            \caption{Snapshots of the particles of the three runs onto the x-y plane. 
            Left-hand panels: the simultaneous minor mergers of compact satellites, run 1A10Bsm.
            Middle panels: the sequential minor mergers of compact satellites, run 1A10Bsq.
            Right-hand panels: the sequential minor mergers of diffuse satellites, run 1A10Csq.}
            \label{snapshot}
        \end{center}
    \end{minipage}
\end{figure*}

Figure~\ref{snapshot} shows five snapshots for three runs, 1A10Bsm, 1A10Bsq and 1A10Csq (Table \ref{table:simulation60_new}).
Elapsed time from the beginning of each run is written in the box of each snapshot.
As described in \S\ref{sec:nbodysimulation}, we stop runs after $t = 3.6$ Gyr of evolution in the minor merger simulations, and analyze the merger remnants at $t = 3.5 - 3.6$Gyr, to allow a small number of surviving satellites to be among the remnants.

In this final state, there are the several surviving cores of the satellite stellar systems among the remnants of the 10 minor mergers.
However, most of stars and the dark matter component of the satellite galaxies have already been stripped or merged into the primary galaxies.
We show that these surviving cores have only a small effect on properties of the merger remnants in \S\ref{sec:surf_stellar}.
In the major merger (run 2A), the first crossing of galaxies is at $\sim 0.5$Gyr, and the merger appears to be completed by $2.3$Gyr.

\subsection{Density profile}\label{sec:prof_stellar}

One of the most revealing properties of the remnants is their density profiles.
Fig. \ref{fig:Rlog_sRholog2} shows the angle-averaged stellar density profiles of the merger remnants for five runs.
In this figure, we show three types of final density profiles for each run: the profile of the stars that initially belong to the primary galaxy (we call ``primary stars''), that of the stars that initially belong to the satellite galaxies (we call ``satellite stars''), and that of all the stars.
For comparison, the density profile of the initial galaxy model is also shown.
The density profiles for all the stars differs significantly between each run. 
Inside the inner region of the half mass radii (summarized in Table \ref{table:simulation60_new}), the gravity of the stellar mass is closely related with the velocity dispersion of the stars in this region, as shown in \S\ref{sec:sigma_profile}.
Below, we discuss these differences in each of the inner and outer sections of the remnants.

\subsubsection{Inner region of the stellar system}\label{sec:inner_stellar}
\begin{figure*}
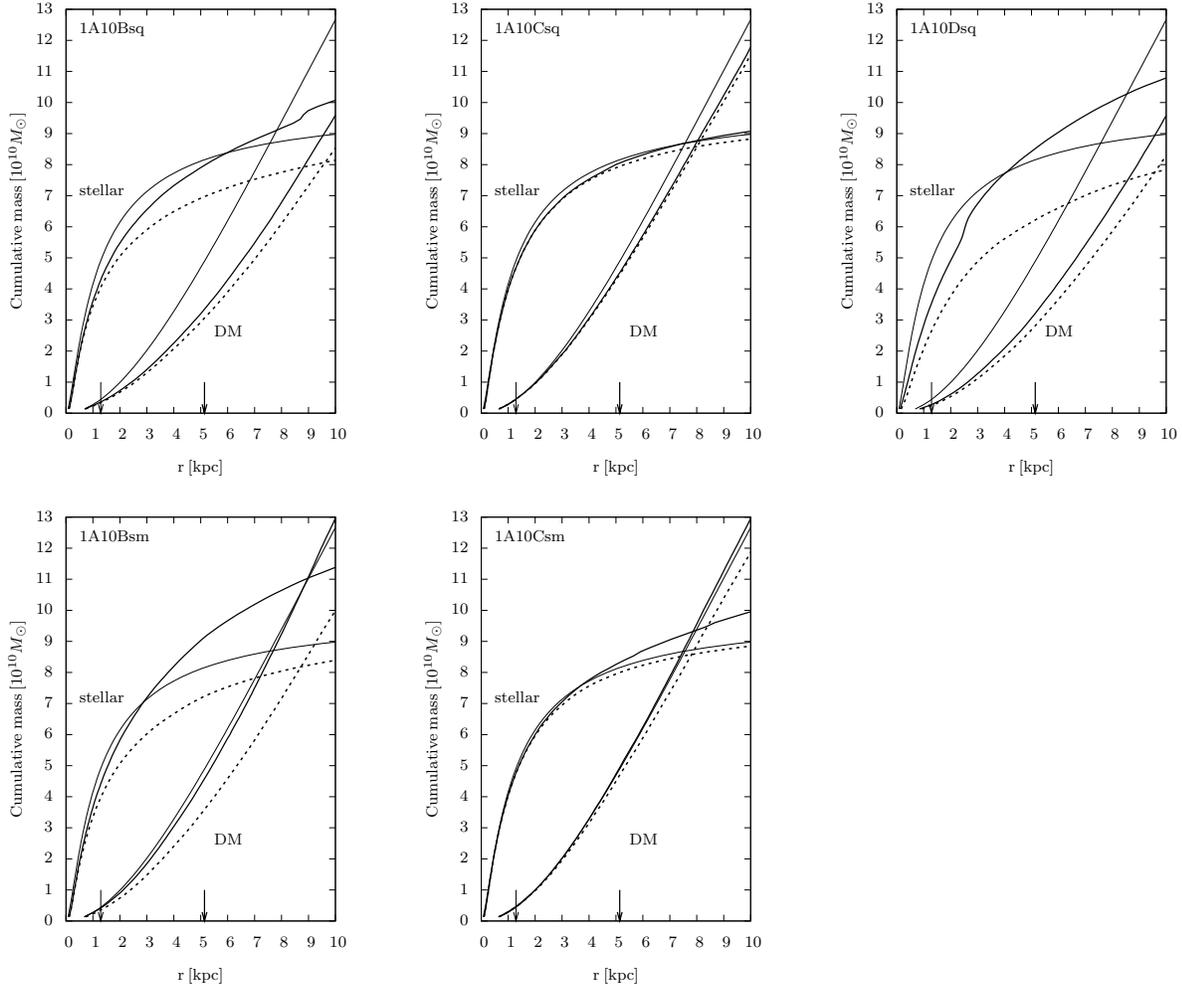

    \begin{minipage}{180mm}
        \begin{center}
            \begin{tabular}{ccc}
                \hspace{-20mm}
                \resizebox{95mm}{!}{\input{./picture/Rlinear_Mlinear_add5_10_high.tex}} &
                \hspace{-45mm}
                \resizebox{95mm}{!}{\input{./picture/Rlinear_Mlinear_add5_10_normal.tex}} &
                \hspace{-45mm}
                \resizebox{95mm}{!}{\input{./picture/Rlinear_Mlinear_1A10cBsq.tex}} \\
                \hspace{-20mm}
                \resizebox{95mm}{!}{\input{./picture/Rlinear_Mlinear_multi_10_5_high.tex}} &
                \hspace{-45mm}
                \resizebox{95mm}{!}{\input{./picture/Rlinear_Mlinear_multi_10_5_normal.tex}} \\
            \end{tabular}
            \caption{
            Cumulative mass distribution of the remnant stellar systems and dark matter particles.
            From top left to bottom right: 1A10Bsq, 1A10Csq, 1A10Dsq, 1A10Bsm, and 1A10Csm.
            Thick solid lines are the cumulative mass distributions of all stars or all dark matter particles of the merger remnants,
            dotted lines are of the primary stars or the primary dark particles.
            Thin lines are of the initial galaxy model.            
            Thick arrows are half-mass radii of the stellar system of the remnants.
            Thin arrows are half-mass radii of the stellar system of the initial primary galaxy model.
            }
            \label{fig:Rlinear_sMlinear}
        \end{center}
    \end{minipage}
\end{figure*}

To make clear the difference between the five runs in the inner regions, we show the cumulative mass distributions of the star particles of the remnants in Fig. \ref{fig:Rlinear_sMlinear}.
We first focus the distributions of the primary stars.
Comparing the distribution of the primary stars of the initial state with that of the final state for the minor mergers of the compact satellites (run 1A10Bsm, 1A10Bsq, and 1A10Dsq), we show that the decrease in the density profiles is due to expansion of the primary stars.
Such expansion is caused by dynamical friction heating from the satellites (\citealt*{2001ApJ...560..636E}, \citealt{2004PhRvL..93b1301M}).
We see in Figure~\ref{fig:Rlinear_sMlinear} that this expansion is smaller for the minor mergers with the diffuse satellites (run 1A10Csm and 1A10Csq).
As shown in Fig. \ref{snapshot}, the compact satellites can survive for a longer time-scale than the diffuse satellites in the tidal field of the primary galaxy  (see also \citealt{2007MNRAS.374.1227B}). This increased lifetime means that the dynamical friction heating can be more effective in expanding the primary stars.
Therefore,the compact satellites are more efficient at heating than the diffuse satellites.

In the inner region, the density profiles of the satellite stars in the remnants depend on the compactness of the satellites.
From Fig. \ref{fig:Rlog_sRholog2} for the compact satellites, we can see that there are more satellite stars in the half mass radii than for the diffuse satellites.
Once again, this is due to the compact satellites surviving against the tidal field of the primary galaxy to reach its inner regions.
For the minor mergers with the diffuse satellites, the satellite stars have a smaller contribution to the density profiles in the half mass radii.
This is because the diffuse satellites are easily disrupted by tidal stripping and tidal shocking from the primary galaxy.
This result is consistent with the minor merger simulations of a single satellite galaxy by \citet{2007MNRAS.374.1227B}.

Compared with the sequential minor mergers, simultaneous minor mergers find a higher number of satellite stars in the half mass radii.
It is known that in a time-varying self-gravitational potential, particles exchange their energies with each other, altering the energy distribution of the particles.
This process is known as violent relaxation (\citealt{1967MNRAS.136..101L}) and produces tightly bound particles.
\S\ref{sec:sigma_profile} will show in more detail that the satellite stars of the simultaneous minor mergers have experienced this process. 
We will also show in the case of sequential minor mergers, that the  violent relaxation process is less effective, and the number of tightly bound particles is smaller than for the simultaneous minor mergers in \S\ref{sec:sigma_profile}.

\subsubsection{Outer region of the stellar system}\label{sec:outer_stellar}
\begin{figure*}
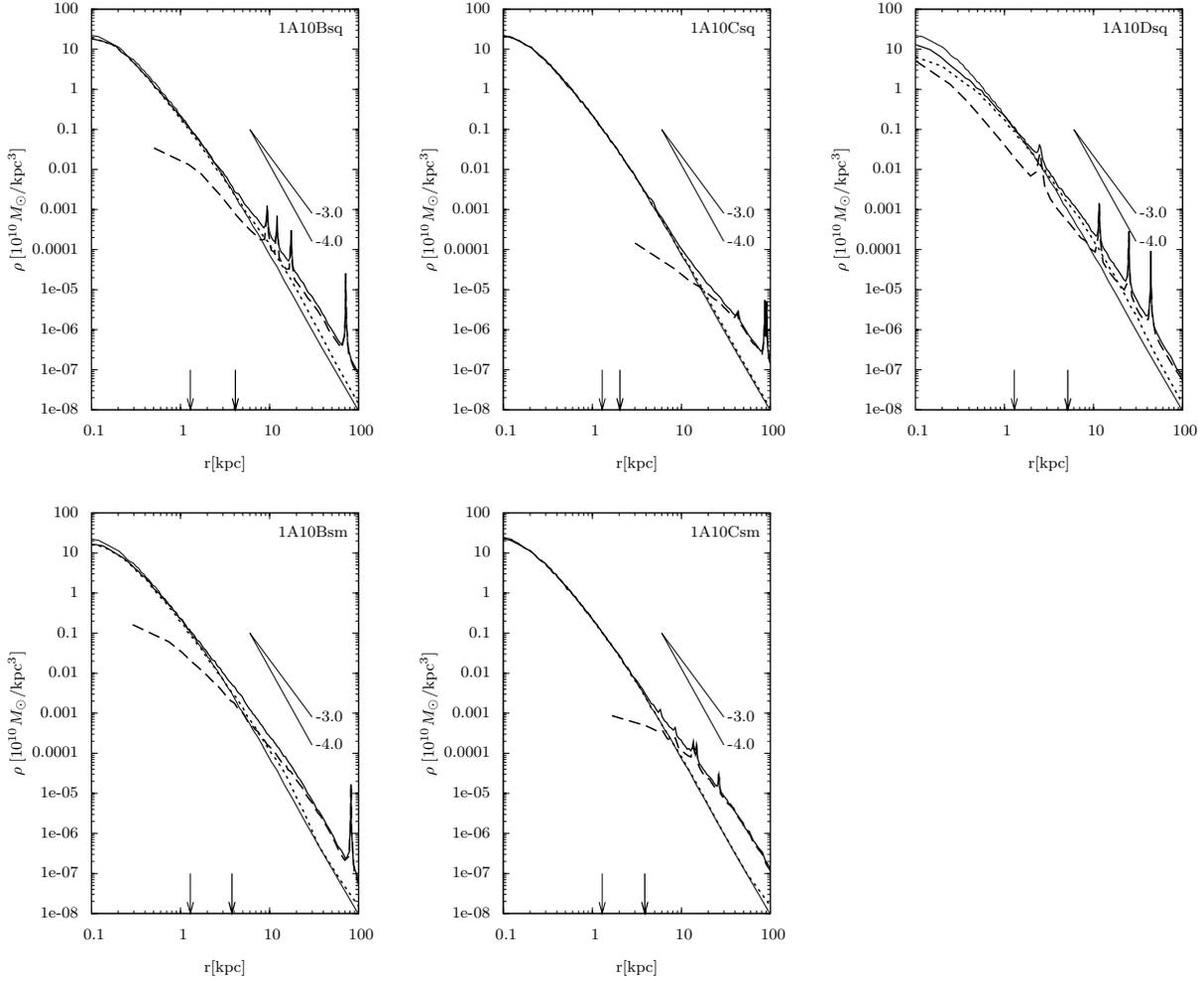

    \begin{minipage}{180mm}
        \begin{center}
            \begin{tabular}{ccc}
                \hspace{-20mm}
                \resizebox{95mm}{!}{\input{./picture/Rlog_sRholog_add5_10_high_6.tex}} &
                \hspace{-45mm}
                \resizebox{95mm}{!}{\input{./picture/Rlog_sRholog_add5_10_normal_6.tex}} &
                \hspace{-45mm}
                \resizebox{95mm}{!}{\input{./picture/Rlog_sRholog_1A10cBsq_6.tex}} \\
                \hspace{-20mm}
                \resizebox{95mm}{!}{\input{./picture/Rlog_sRholog_multi_10_5_high_6.tex}} &
                \hspace{-45mm}
                \resizebox{95mm}{!}{\input{./picture/Rlog_sRholog_multi_10_5_normal_6.tex}} &
                \hspace{-45mm}
            \end{tabular}
            \caption{The angle-averaged density profiles of the remnant stellar systems of runs. 
            From top left to bottom right: 1A10Bsq, 1A10Csq, 1A10Dsq, 1A10Bsm, and 1A10Csm. 
            Thick solid lines are the density profiles of all stars of the merger remnants, dotted lines are of the stars that belonged to the primary galaxy (we call primary stars), and dashed lines are of the stars that belonged to the satellite galaxies (we call satellite stars).
            Thin lines are of the initial galaxy model.
            The slops of $r^{-3.0}$ and $r^{-4.0}$ are also shown for comparison.
            Thick arrows are half-mass radii of the stellar system of the remnants.
            Thin arrows are half-mass radii of the stellar system of the initial primary galaxy model.
            }
            \label{fig:Rlog_sRholog2}
        \end{center}
    \end{minipage}
\end{figure*}

Fig. \ref{fig:Rlog_sRholog2} shows that the increase in density in the outer regions of the density profile is dominated by the accretion of the satellite stars onto the outside of the half mass radii.
The density profiles of the primary stars is hardly changed, as shown in Fig. \ref{fig:Rlog_sRholog2} and follows a $r^{-4}$ profile in the outer region.
On the other hand, the slope of the density profiles of the satellite stars follows an $r^{-3}$ up to several 100 kpc.
Then for radii larger than the several 100 kpc, the slope declines as $\rho \propto r^{-4}$.
These difference show that the two components have distinct origins; an already formed system and a newly accreted system.
In particular, in the case of the sequential minor mergers of the diffuse satellites (run 1A10Csq), the satellite stars contribute to the envelope of the density profile of the remnant as an excess component, as shown in Fig. \ref{fig:Rlog_sRholog2}.
Similar results are obtained by \citet*{2006MNRAS.365..747A}.
\citet{2006MNRAS.365..747A} simulate an isolated disc galaxy with cosmological initial conditions
and show that the density profile of the accreted stellar component follows the slope $\rho \propto r^{-3}$ near the edge of the primary stellar component of the galaxy.
In addition, the slope reaches $\rho \propto r^{-4}$ or steeper near the virial radius of the system.
Several recent observations of compact massive ETGs at low-z ($z<0.5$) show that the surface brightness profiles of the ETGs are well fitted with two-component S\'{e}rsic profiles (\citealt*{2010ApJ...709L..58S}, \citealt{2011ApJ...733...45S}).
While these S\'{e}rsic indexes are relatively small (n$\la2$), these galaxies may have experienced the size growth by minor mergers like our simulations.

\subsubsection{Dark matter halo}
Fig. \ref{fig:Rlinear_sMlinear} also shows cumulative mass distributions of the dark matter particles of the merger remnants of the five runs.
The mass of dark matter haloes are smaller than the stellar mass in the inner regions.
However, change of dark matter mass distribution is important, since this affects the velocity dispersion of the stars in this region, as shown in \S\ref{sec:sigma_profile}.
In the case of the minor mergers of the compact satellites, the dark matter particles that initially belong to the primary galaxy (we call ``primary DMs'') expand due to dynamical friction heating.
Compared with the simultaneous minor mergers, sequential minor mergers have only small number of dark matter particles that initially belong to the satellite galaxies (we call ``satellite DMs'') contribute to the density profiles within 10kpc from the centre of the merger remnant.
For simultaneous minor mergers, a significant amount of the satellites' DM is deposited in the central 10kpc.
Thus, sequential minor mergers of compact satellites are the most efficient in expanding of the dark matter distribution and in decreasing of the central dark matter density.

\subsection{Stellar surface density profile}\label{sec:surf_stellar}

\begin{figure}
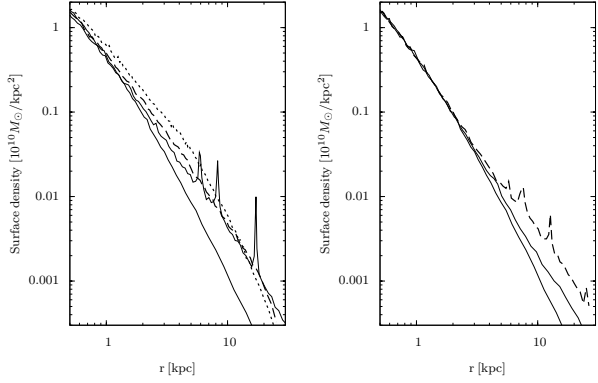

    \begin{minipage}{84mm}
        \begin{center}
            \begin{tabular}{cc}
                \hspace{-23mm}
                \resizebox{76mm}{!}{\input{./picture/Rlog_sSigmalog_1_5.tex}} &
                \hspace{-40mm}
                \resizebox{76mm}{!}{\input{./picture/Rlog_sSigmalog_2_5.tex}} \\
            \end{tabular}
            \caption{Left: Angle-averaged surface density profiles of the remnant stellar systems in one projection, 2A(dotted), 1A10Bsm(dashed), 1A10Bsq(thick solid), and initial model A (thin solid).
            Right: The same as the left panel, but for 1A10Csm (dashed), 1A10Csq(thick solid), and initial model A (thin solid).
            }
            \label{fig:Rlinear_sSigmalog}
        \end{center}
    \end{minipage}
\end{figure}

We show the angle-averaged stellar surface density profiles of the merger remnants in Fig. \ref{fig:Rlinear_sSigmalog}.
For the evolution in the surface density profiles, the choice of projection direction does not significantly alter the result.
For the major merger (run 2A), the surface densities in $r\la 1$kpc increases from the initial state.
On the other hand, for the minor mergers, the surface densities are approximately stationary in $r\la1$kpc.
In $\ga2$kpc, the growth in the surface densities is dominated by the accretion of the satellites onto the outer envelope.
The latter change agrees well with the recent observational results from \citet{2010ApJ...709.1018V} and is an example of ``inside-out growth''.
These profiles depend on the compactness of the satellite galaxies.
In Fig. \ref{fig:Rlinear_sSigmalog}, we show that inside-out growth is significant when the satellite galaxies are diffuse.

In the case of 10 sequential minor mergers, there are several surviving cores of the satellite stellar systems within $r_{\mathrm{trunc}}$ of the remnants.
Due to these cores, the remnant sizes might be artificially large in our analysis.
We examine effects of these cores in determining sizes and mass of the merger remnants by removing the satellite cores from the surface density profile of run 1A10Bsq.
We find that the remnant after the removing is $\sim 0.67\times 10^{10} M_{\odot}$ less massive and $\sim 0.32$kpc smaller than with the cores included.
This size change is much smaller than the size growth of the remnant of 1A10Bsq that increased from 1kpc to 3kpc.
Therefore, in our analysis, these cores do not affect the size growth of the remnants significantly.

\subsubsection{S\'{e}rsic profile fitting}
In order to compare the structures of our merger remnants with observational results quantitatively,
we fit the stellar surface density profiles of the remnants along the major and the minor axes derived by the method in \S3.5 with the \citet{1968adga.book.....S} $R^{1/m}$ law:
\begin{equation}
  I(R) = I_0 \exp \left[ -b(m) \left( \frac{R}{R_e} \right)^{1/m} \right],
\end{equation}
where $b(m) \sim 2m - 1/3 + 4/(405m)$ (\citealt{1999A&A...352..447C}).
We choose the radial range [0.1, 10] kpc for this fitting.
In this range, the surface density profiles are well fitted with a single S\'{e}rsic profile.
We note that the S\'{e}rsic index $m$ is very sensitive to the fitting range (see also \citealt{2005MNRAS.362..184B}).

We find that the S\'{e}rsic indexes of the remnants increase from our initial model in all runs.
These values are consistent with the observed massive ETGs (e.g. \citealt*{2002A&A...386..149B}, \citealt{2003AJ....125.2936G}).
In particular, for run 1A10Bsq, the S\'{e}rsic index increases significantly because of the marked inside-out growth.
On the other hand, for run 1A10Csq, the S\'{e}rsic index increases less.
As described above, this is because that the stars of diffuse satellites are easily stripped by the tidal field of the primary galaxies, and few satellite stars are added inside of 10kpc from the centre of the remnant.
The increase of the S\'{e}rsic index significantly depends on the compactness of the satellites and the manner of minor mergers (sequential or simultaneous).

\begin{table}
  \centering
  \begin{minipage}[t]{80mm}
    \caption{The S\'{e}rsic index. The $b-c$ plane means the projection along the major axis.
    The remnant is projected in the plane with the intermediate axis $b$ and the minor axis $c$.
    The $a-b$ plane is the projection along the minor axis.
    This plane contains the major axis $a$ and the intermediate axis $b$.
     }
    \label{table:sersic}
    \begin{tabular}{@{}ccc@{}}
      \hline
      \hline
      Name & $b-c$ plane & $a-b$ plane  \\

      \hline
      A (initial)    & 4.07 & 4.04  \\
      \hline
      2A               & 5.00 & 6.16  \\
      1A10Bsm  & 6.51 & 6.32  \\
      1A10Bsq   & 11.3 & 8.47  \\
      1A10Csm  & 8.41 & 8.27  \\
      1A10Csq   & 4.83 & 4.76  \\
      1A10Dsq   & 4.66 & 7.87  \\
      \hline
    \end{tabular}
  \end{minipage}
\end{table}

\subsection{Velocity dispersion profile}\label{sec:sigma_profile}
In Figure~\ref{fig:Rlog_Vdisplinear} we show the evolution of the line-of-sight stellar velocity dispersion profiles for the four runs with minor mergers.
As with the surface density profiles,  the choice of projection does not affect our main results.
Fig. \ref{fig:Rlog_Vdisplinear} shows that in the cases of minor mergers with the compact satellites, the velocity dispersion of the stellar systems in the central region decreases from the initial model more strongly than for the minor mergers with the diffuse satellites.
This is because the central region is gravitationally heated by the compact satellites.
We compare the velocity dispersion of the remnant of  run 1A10Bsm with that of run 1A10Bsq at $\sim 2.4$kpc, which is near the effective radii of these remnants.
It is obvious from Figure~\ref{fig:Rlog_Vdisplinear} that for run 1A10Bsq, the velocity dispersion in this radius is smaller than for run 1A10Bsm.
This is because for run 1A10Bsq, few satellite stars and satellite DMs are deposited within this radius.
For run 1A10Bsm, relatively large amounts of the satellite stars and satellite DMs are deposited within this radius, and 
the velocity dispersion of the remnant increases from the initial value in this region.
Therefore, in the case of the sequential minor mergers of the compact satellites, the velocity dispersion is smaller in this region.

\begin{figure}
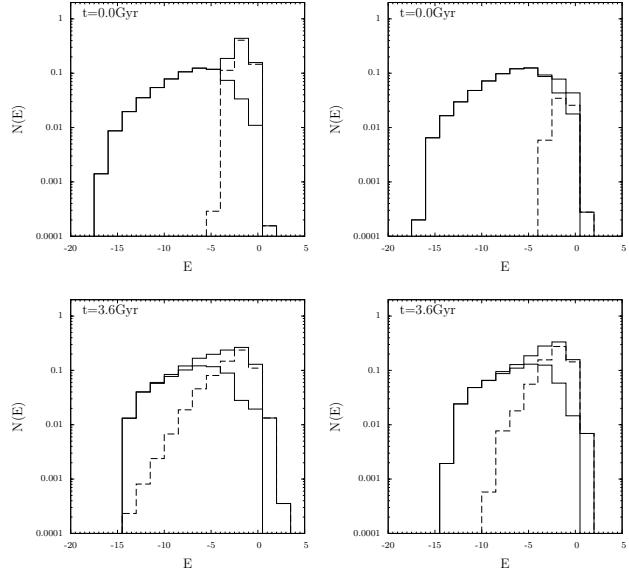

    \begin{minipage}{84mm}
        \begin{center}
            \begin{tabular}{cc}
                \hspace{-10mm}
                \resizebox{55mm}{!}{\input{./picture/diff_e_dist_star_multi5_high_initial.tex}} &
                \hspace{-18mm}
                \resizebox{55mm}{!}{\input{./picture/diff_e_dist_star_add5_high_initial.tex}} \\
                \hspace{-10mm}
                \resizebox{55mm}{!}{\input{./picture/diff_e_dist_star_multi5_high_final.tex}} &
                \hspace{-18mm}
                \resizebox{55mm}{!}{\input{./picture/diff_e_dist_star_add5_high_final.tex}} \\
            \end{tabular}
            \caption{Left-hand panels: initial (top panel) and final (bottom panel) differential energy distributions of the star particles for the simultaneous minor mergers of compact satellites (run 1A10Bsm).
            The thick solid, thin solid, and dashed line refer to the distributions of total stars, primary stars, and satellite stars, respectively.
            The energy per unit mass $E$ is normalized to the internal unit of our simulations.
            Right-hand panels: same as left-hand panels, but for the sequential minor mergers of compact satellites (run 1A10Bsq).}
            \label{n_e_evolve}
        \end{center}
    \end{minipage}
\end{figure}

To interpret the difference in the mass deposited during both runs, we show the differential energy distributions $N(E)$ for the  initial and final stellar components in Fig. \ref{n_e_evolve}, where $N(E)dE$ represents the number of particles of specific energy between $E$ and $E+dE$.
For the simultaneous minor merger case, the distribution of the satellite stars broadens during the mergers and the low energy stars are more bound than the initial state.
This broadening of $N(E)$ is similar to the behavior of $N(E)$ of cold collapse simulations (\citealt{2008gady.book.....B}).
This means that in this case the violent relaxation is effective for the satellite stars.
Due to such evolution, the satellite stars are more bound and strongly invade the centre of the primary galaxy.
On the other hand, there is smaller change of $N(E)$ for the sequential minor merger case.
This means that in this cases the violent relaxation is not effective and the satellite stars can not invade the inner region.
In both simulations shown in Fig. \ref{n_e_evolve}, the number of the most bound primary stars are reduced.
This unbinding is induced by the dynamical friction heating during the minor mergers.

We also investigate anisotropy of the velocity dispersions of the stellar merger remnants.
We measure an anisotropy parameter for the remnants
\begin{equation}
\beta = 1- \left( \frac{\sigma_{\theta}^2}{\sigma_{r}^2} \right)
\end{equation}
where $\sigma_{\theta}$ and $\sigma_{r}$ are the azimuthal and radial velocity dispersions with respect to the centre of the remnants, respectively.
Here, we assume $\sigma_{\theta} = \sigma_{\phi}$.
In the bottom panels of Fig. \ref{fig:Rlog_Vdisplinear}, we show the anisotropy of the merger remnants.
For all runs, radial anisotropy increases with radii.
It is obvious that in the case of sequential minor mergers, radial anisotropy is smaller than that for simultaneous minor mergers.
This result is consistent with \citet{2005AIPC..804..333A} who shows that radial anisotropy becomes smaller for larger time intervals between mergers in sequential galaxy mergers in galaxy groups.
We note that for the sequential minor mergers satellite orbits are radial with respect to the `initial' position of the primary galaxy, and actually these orbits may be more circular than for the simultaneous minor mergers.
This may affect the small radial anisotropy for the sequential minor mergers.
Our results suggest that the manner of minor mergers influences the radial anisotropy of the merger remnants.

\subsection{Evolution of stellar mass, $R_e$, and $\sigma_e$}\label{sec:r_e_sigma_e}
We summarize the results for stellar masses, the effective radii $R_e$, and the velocity dispersions $\sigma_e$ of the merger remnants in Table \ref{table:simulation60_new}.
The minor mergers of the compact satellites are more efficient in increasing the mass of the remnants than those of the diffuse satellites.
More than half of the satellite stars in the compact satellites contribute to the mass of the remnants.
On the other hand, in the case of the diffuse satellites (model C), the satellites are disrupted by tidal stripping and shocking during the merging, and most of satellite stars of the diffuse satellites are distributed in the envelope of the remnants.
These are too diffuse to observed and do not contribute to the mass of the remnants (run 1A10Csq).

Changes of $R_e$ and $\sigma_e$ are influenced both by the manner of the minor mergers and by the compactness of the satellite galaxies.
For the minor mergers of the compact satellites, the sequential cases cause larger size growth and a greater decrease in the velocity dispersion compared to that in the simultaneous cases.
On the other hand, for the minor mergers with the diffuse satellites, the simultaneous cases cause larger size growth, while the decrease in the velocity dispersion is weak in both the sequential and the simultaneous cases.
There are two processes which are closely related with the changes of $R_e$ and $\sigma_e$.
The first process is the expansion of the primary stars by dynamical friction heating.
This increases the size of the primary galaxy and decreases the velocity dispersion, if it is principally determined by the mass distribution of the primary stars.
In the case of minor mergers with the compact satellites, this expansion is more effective in both the sequential and the simultaneous cases than for the minor mergers with the diffuse satellites (Fig. \ref{fig:Rlinear_sMlinear}).
The second process is mass deposit of the satellite stars by the minor mergers.
The mass deposit increases the size of the remnant efficiently for all runs.
It is important for change to the velocity dispersion whether the satellite stars are deposited within the effective radii of the merger remnants or beyond it.
If a significant number of the satellite stars are deposited within the effective radii, the velocity dispersions of the remnants increase.
In the case of the sequential minor mergers, the mass deposit within the effective radii is small.
In this case, the velocity dispersion is mainly determined by the expansion of the primary stars, and the velocity dispersion decreases.
These two processes also cause a changes in the density and the velocity dispersion profiles (see \S\ref{sec:prof_stellar}, \S\ref{sec:sigma_profile}).
We note that comparison of run 1A10Bsq, 1A10Csq, and 1A10Dsq shows that for the minor mergers of the compact satellites, the effective radii increases more than for the minor mergers of the diffuse satellites.
This result is apparently contrary to expectations based on the virial theorem and the energy conservation (e.g. \citealt{2009ApJ...699L.178N}).
However, because of tidal disruption of the diffuse satellites as described above, the size and mass increase is weak in the case of the diffuse satellites.

We present two important relations from our results.
The first is the relation between the size growth and the mass growth.
The second is the relation between the decrease in velocity dispersion and the mass growth.
We parameterise the relation between the size growth and the mass growth as 

\begin{equation}
\frac{R_{e,f}}{R_{e,i}} = \left( \frac{M_{f}}{M_{i}} \right)^{\alpha} \label{eq:size_growth}
\end{equation}
to quantify the efficiency of the size growth (hereafter we call $\alpha$ the size growth efficiency), where $R_{e,i}$ and $R_{e,f}$ are a initial and a final effective radius, $M_{i}$ and $M_{f}$ are a initial and a remnant stellar mass, respectively.
We also parameterise the relation between the change of the velocity dispersion and the stellar mass growth as 

\begin{equation}
\frac{\sigma_{e,f}}{\sigma_{e,i}} = \left( \frac{M_{f}}{M_{i}} \right)^{-\gamma}
\end{equation}
to quantify the efficiency in the change in the velocity dispersion.
Table \ref{table:simulation60_new} shows $\alpha$ and $\gamma$ for each run.
As shown in Table \ref{table:simulation60_new}, the cases of the sequential minor mergers, $\alpha$ and $\gamma$ are generally higher than that for the simultaneous minor mergers.
In particular, we show that the sequential minor mergers of model B are the most efficient in the size growth, $\alpha \simeq 2.7$.
$\gamma$ is larger for runs with model D, the most compact satellite.
In the case of the sequential minor mergers of the diffuse satellites (run 1A10Csq), $\alpha$ and $\gamma$ are also larger.
In this case, only the expansion is effective.
However, here the satellite stars do not contribute to the mass increase of the merger remnant because of the tidal disruption of the diffuse satellites as described above.
Therefore, the minor mergers of the diffuse satellites do not contribute the size growth significantly.

As described in the previous section, in the case of 10 sequential minor mergers, there are several surviving cores from the satellite stellar systems in the remnants within $r_{\mathrm{trunc}}$.
Using the prescription for removing satellite cores for 1A10Bsq, we find that the size growth efficiency $\alpha \simeq 2.7$ with or without the satellites core.
Therefore in our analysis, these cores do not affect the size growth efficiency of the remnants significantly.

\section{Discussion}

\subsection{Size growth efficiency}
As shown in Table \ref{table:simulation60_new}, the size growth efficiencies $\alpha$ are higher than for an analytical estimation (e.g. \citealt{2009ApJ...699L.178N}) or found numerically in previous works (e.g. \citealt{2006MNRAS.369.1081B}).
We check these high efficiencies using a simple analytical argument:
Under the assumptions of energy conservation, single component galaxies, and a parabolic orbit in a minor merger, \citet{2009ApJ...699L.178N} show that the ratio of the final to initial gravitational radius is

\begin{equation}
\frac{r_{g,f}}{r_{g,i}} = \frac{(1+\eta)^2} {(1+\eta \epsilon)}. \label{eq:size_ratio}
\end{equation}
Here, we define $\eta \equiv M_2 / M_1$ and $\epsilon \equiv \langle v_{1}^2 \rangle / \langle v_{2}^2 \rangle$, where $M_1$ and $\langle v_{1}^2 \rangle$ are the mass and the velocity dispersion of the more massive galaxy before the minor merger $M_2$ and $\langle v_{2}^2 \rangle$ are those of the less massive galaxy.
This formula is also valid for two-component galaxy models if the dark matter halo profile strictly follows the stellar density profile (\citealt{2012MNRAS.tmp.2680N}). 
We adopt this assumption in the following discussion.
$r_g$ in Eqn. (\ref{eq:size_ratio}) can be replaced with $R_e$, where $r_g$ and $R_e$ satisfy the following equation,

\begin{equation}
E \equiv -f \frac{GM_{*}^{2}} {R_{e}} = - \frac{1}{2} \frac{GM_{*}^2} {r_g},
\end{equation}
where $E$ is the energy of the stellar system and $f$ is a structural parameter that depends on the stellar and dark matter density profiles.
We obtain

\begin{equation}
\frac{R_{e,f}}{R_{e,i}} = \frac{f_f}{f_1} \frac{(1+\eta)^2} {(1+\eta \epsilon)}, \label{eq:size_ratio_re}
\end{equation}
where $f_1$ and $f_f$ are the structural parameters of the progenitor and remnant stellar systems.
Here, we assume that two progenitors have the same structural parameter $f_1$.
This assumption is valid for runs 1A1B and 1A1C.
Thus, if $f$ increases in a dry minor merger, $R_e$ increases more than for the case where $f_f = f_1$.
From Eqn. (\ref{eq:size_ratio_re}), the size growth efficiency can be written as

\begin{equation}
\alpha = \frac{d\ln R_e} {d\ln M_{*}} = \frac{\ln (f_f / f_1)} {\ln (1+\eta)} + 2 - \frac{\ln (1+\eta \epsilon)} {(1+\eta)}. \label{eq:alpha_formula}
\end{equation}
We estimate $\alpha$ from $f_f / f_1$, $\eta$, and $\epsilon$ from our numerical results with Eqn. (\ref{eq:alpha_formula}).
In the case of run 1A1B $\alpha = 2.2$, since $f_f / f_1 = 1.07$, $\eta= 0.1$, and $\epsilon=0.446$.
For run 1A1C $\alpha = 2.3$, since $f_f / f_1 = 1.06$, $\eta= 0.1$, and $\epsilon=0.297$.
For run 1A1D $\alpha = 2.0$, since $f_f / f_1 = 1.06$, $\eta= 0.1$, and $\epsilon=0.580$.
We analyse the merger remnants of run 1A1B, 1A1C, and 1A1D (see Table \ref{table:strucrural}), defining the merger remnants as bound particles that are consistent with the energy conservation argument here.
We obtain size growth efficiencies of $\alpha = 2.4$ (run 1A1B), and $\alpha = 2.2$ (run 1A1C).
These values agree well with the analytical argument.
On the other hand, in the case of run 1A1D, $\alpha = 2.5$ which does not agree well with the analytical result.
In this case, the change in the stellar and dark matter halo profiles is larger due to the strong dynamical friction heating.
Thus, this disagreement may stem from the assumption in the analytical argument that the dark matter halo profile follows the stellar density profile, which is failing to hold in this case.
Alternatively, the disagreement may be because the assumption that the two progenitors have the same structural parameter $f_1$ is invalid in this case.
Therefore, we suggest that our high size growth efficiency can be interpreted as the increase of structural parameters $f$ of merger remnants (see also a similar analysis in \citet{2006MNRAS.369.1081B}).
Similar results of the high size growth efficiency for minor mergers on radial orbits are shown in \citet{2012MNRAS.tmp.2680N}.

\begin{table}
  \centering
  \begin{minipage}[t]{80mm}
    \caption{Physical properties of progenitors and merger remnants in binary minor merger simulations.
    Note that we define the merger remnants as bound particles here.
    Each column is similar to Table \ref{table:simulation60_new} except for fifth column, structure parameter of the galaxy.
     }
    \label{table:strucrural}
    \begin{tabular}{@{}ccccc@{}}
      \hline
      \hline
      Name & stellar mass  & $R_e$ & $\alpha$ & $f$  \\
                  &  ($M_{\odot}$) & (kpc)  &   &  \\ 

      \hline
      modelA  & $ 9.97 \times10^{10}$ & 0.995 & & 0.173 \\
      modelB  & $ 0.996 \times10^{10}$ & 0.453 & & 0.177 \\ 
      modelC  & $0.994 \times10^{10}$ & 1.11 & & 0.173 \\
      modelD  & $0.999 \times10^{10}$ & 0.261 & & 0.163 \\
      1A1B      & $10.8\times10^{10}$ & 1.21 & 2.40 & 0.185 \\
      1A1C      & $10.7\times10^{10}$ & 1.16 & 2.20 & 0.183 \\
      1A1D      & $10.9\times10^{10}$ & 1.23 & 2.54 & 0.183 \\
      \hline
    \end{tabular}
  \end{minipage}
\end{table}

\subsection{Comparison with observations}
Our main focus is whether the dry minor mergers can explain the size evolution problem of ETGs and reproduce their local scaling relations, in particular, the stellar mass-size and the stellar mass-velocity dispersion relation.
We compare our numerical results with these results here.

\subsubsection{Stellar mass-size relation}
\begin{figure}
  \includegraphics[width=84mm]{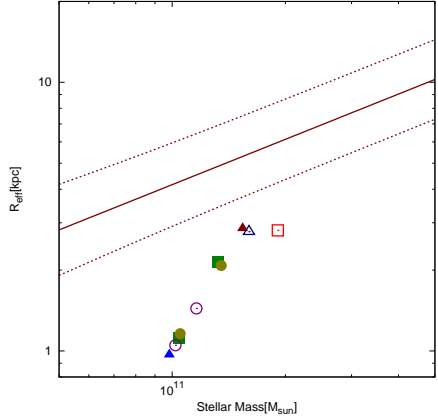}
  \caption{Stellar mass-size relation of the merger remnants and the initial galaxy model. 
  Solid triangle is for the initial galaxy model A. 
  Empty square is for the run 2A.
  Solid square is for the run 1A10Bsm. 
  Empty circle is for the run 1A10Bsq.
  Filled circle is for the run 1A10Csm.
  Empty triangle is for the run 1A10Csq.
  The solid line shows the observed stellar mass-size relation for ETGs (\citealt{2003MNRAS.343..978S}) with the dispersion by the dotted lines.
  } 
  \label{fig:M_R}
\end{figure}

In Fig. \ref{fig:M_R}, we compare our numerical results with the local stellar mass-size relation given by the SDSS (\citealt{2003MNRAS.343..978S}).
We show that the sequential minor mergers of the compact satellites cause the most efficient size growth, while all results move the compact massive ETGs toward the local relation in Fig. \ref{fig:M_R}.
Interestingly, the sequential minor mergers have $\alpha \simeq 2.7$ (Table \ref{table:simulation60_new}), which is much larger value than that obtained via the simple analysis based on the virial relation, $\alpha \sim 2$ (\citealt{2009ApJ...697.1290B}, \citealt{2009ApJ...699L.178N}), with the assumptions of energy conservation, single component galaxies, and parabolic orbit of minor mergers.
The sequential minor mergers of compact satellites can evolve a compact massive ETGs to that with the the average size of ETGs in the local Universe by a factor of $\sim2$ mass growth.
As shown in \S2, such mass growth is likely to occur for the majority of the high-z massive galaxies.

Our high growth efficiency is consistent with the observational constraint by \citet{2009ApJ...697.1290B}.
In \citet{2009ApJ...697.1290B}, they suggest that the size growth efficiency, $\alpha$, must be $\alpha \ga 2$ in order to satisfy that the number density of descendants of the high-z compact massive ETGs does not exceed constraints imposed by the z = 0 galaxy mass function.
Our result for the size growth efficiency for the sequential minor mergers, $\alpha \simeq 2.7$, is consistent with their results.

\citet{2012ApJ...746..162N} estimate the size evolution of ETGs using the CANDELS survey and conclude that observed rapid size growth from $z \simeq 2$ to $z \simeq 1$ cannot be explained by the dry minor merger scenario, if they assume $\alpha \sim 1.3 -1.6$ motivated by the previous studies (e.g. \citealt{2009ApJ...706L..86N}).
However, our results show a significantly higher size growth efficiency $\alpha \simeq 2.7$ which may invalidate their conclusion.

\subsubsection{Stellar density}
As shown in Table \ref{table:simulation60_new}, we find that the mean stellar densities within 1kpc, $\bar{\rho} (<1\mathrm{kpc})$, decrease by only a small factor in all runs.
On the other hand, for the sequential minor mergers of the compact satellites, the mean stellar density within the effective radius, $\bar{\rho} (<R_e)$, decreases by more than an order of magnitude.
This evolution is consistent with several recent observational studies (\citealt{2009ApJ...697.1290B}, \citealt{2010ApJ...709.1018V}) which show evidence to suggest that the compact massive ETGs at high-z are the central stellar systems in the normal nearby ETGs.
This scenario is called the ``inside-out'' growth of ETGs.
We note that the decrease of $\bar{\rho} (<R_e)$ in our results exceeds that from simple analysis (\citealt{2009ApJ...699L.178N}), corresponding to a size growth larger than that of the simple estimate.

\subsubsection{Stellar mass-velocity dispersion relation}

\begin{figure}
  \includegraphics[width=84mm]{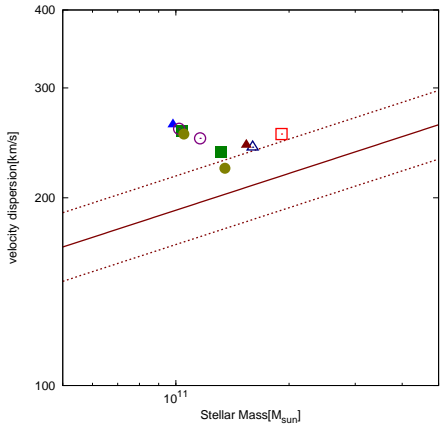}
  \caption{Stellar mass-velocity dispersion relation of the merger remnants and the initial galaxy model.
  Symbols are the same as in Figure \ref{fig:M_R}.
    The solid line shows the observed stellar mass-velocity dispersion relation for ETGs (\citealt{2009ApJ...706L..86N}) with the one-sigma scatter by the dotted lines.
  } 
  \label{fig:sM_Vdisp}
\end{figure}

In Fig. \ref{fig:sM_Vdisp}, we compare our results with the local stellar mass-velocity dispersion relation derived by \citet{2009ApJ...706L..86N}.
We find that the remnant of the sequential minor mergers of the compact satellites lies on the local relation, and the velocity dispersions of all minor mergers decrease toward the local relation.
We note that the decrease in the velocity dispersion for almost all runs is less effective than that derived by simple analysis using the virial relation, $\gamma \sim 1/2$ (\citealt{2009ApJ...699L.178N}).

\subsection{Size growth expected from the Millennium Simulation Database}

As shown in \S 2, we analyze the Millennium Simulation Database given by \citet{2007MNRAS.375....2D} to derive the cumulative mass growth via minor mergers.
In this subsection, we present the size growth for the sample galaxies derived by the Millennium Simulation Database,  using the size growth formula given by Eqn. (\ref{eq:size_growth}) assuming the size growth efficiency $\alpha$.
We confirm that for $\sim 95\%$ ($\sim 91\%$) of our sample galaxies, $\sim 80\%$ ($\sim 90\%$) of the bulge mass at $z=0$ consist of stars that belonged to their progenitor galaxies.
Thus, they mainly increase their masses by dry mergers.
This is mainly because \citet{2007MNRAS.375....2D} use strong supernova and AGN feedback models.
A small fraction of our sample galaxies increases their masses by star formation.
These `wet mergers' may influence their size increase.
In this discussion, we neglect these `wet mergers', since their number is small.
Here, we divide the sample galaxies into two categories: brightest cluster galaxies (BCGs) and normal ETGs.
We define the BCGs as the galaxies that are the central galaxies in FOF haloes with mass of $\ga 10^{14} M_{\odot}$, otherwise we define them as a normal ETGs.

For the sample galaxies, we derive the average size growth factor $\langle R_{e,f}/R_{e,i} \rangle$ by the following procedure:
First, for each sample galaxy we calculate the mass increase through minor mergers from $z=2.07$ to $z=0$, $\Delta m_{\mathrm{minor}}$.
Second, we calculate the mass growth factor $M_{f}/M_{i}$ by the minor mergers from $z=2.07$ to $z=0$, where $M_i$ is the bulge mass at $z=2.07$ and $M_f = M_i + \Delta m_{\mathrm{minor}}$.
Third, we calculate the size growth factor $R_{e,f}/R_{e,i}$ with Eqn. (\ref{eq:size_growth}) for each sample galaxy, where we assume a size growth efficiency $\alpha$.
Here, we adopt the same size growth efficiency for all sample galaxies.
Finally, we average the size growth factors of all sample galaxies as $\langle R_{e,f}/R_{e,i} \rangle$.

\begin{figure}
  \resizebox{84mm}{!}{\input{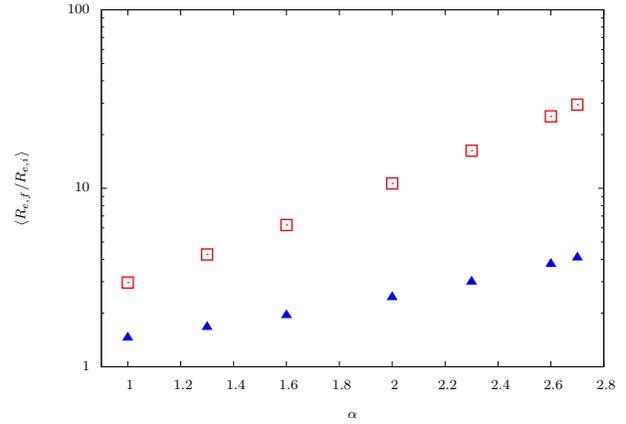}}  
  \caption{The average size growth factor $\langle R_{e,f}/R_{e,i} \rangle$ of our sample galaxies derived by the Millennium Simulation Database for various size growth efficiencies we assume.
  The derivation of $\langle R_{e,f}/R_{e,i} \rangle$ are described in the text.
  Horizontal axis shows the size growth efficiency $\alpha$.
  Empty square is BCGs.
  Solid triangle is normal ETGs.
  } 
  \label{fig:alpha_gf}
\end{figure}

We show the dependence of the average size growth factor on the adopted size growth efficiencies in Fig. \ref{fig:alpha_gf}.
It is obvious that $\alpha \ga 2.3$ is needed to explain the observed size evolution of ETGs from $z\sim2$ to $z=0$ (a factor of three to five in size growth, e.g. \citealt{2006ApJ...650...18T}, \citealt{2007MNRAS.382..109T}, \citealt{2008ApJ...687L..61B}, \citealt{2008A&A...482...21C}, \citealt{2008ApJ...677L...5V}).
This constraint is consistent with the high size growth efficiency derived by our numerical simulations.
On the other hand, the lower size growth efficiency $\alpha = 1.3$ derived by the simultaneous dry minor merger simulations of \citet{2009ApJ...706L..86N} is insufficient to explain the observational results.
Therefore, we suggest that in the dry minor merger scenario, the sequential minor mergers we performed in this paper are responsible for the size evolution of the ETGs.

We should note that we use the maximum size growth efficiency in the above discussion, whereas the choice of $\alpha \simeq 2.7$ corresponds to the maximum predicted size increase and the assumption that all minor mergers are those with mass ratios of $M_{2}/M_{1} < 1/10$ and their orbits are parabolic and head-on.
In fact, there are a wide range of mass ratios for minor mergers in cosmological merger histories.
For example, \citet{2012ApJ...744...63O} find that the typical mass ratio of stellar mergers is 1:5 with cosmological hydrodynamical simulations.
Additionally, there are satellite galaxies which have different properties, e.g. morphology, compactness, and gas content, which may influence the size growth of compact massive ETGs.
In the future studies, we will take into account the mass ratio, orbits, and the many satellite properties for more robust predictions based on the dry minor merger scenario.

\begin{figure}
  \resizebox{84mm}{!}{\input{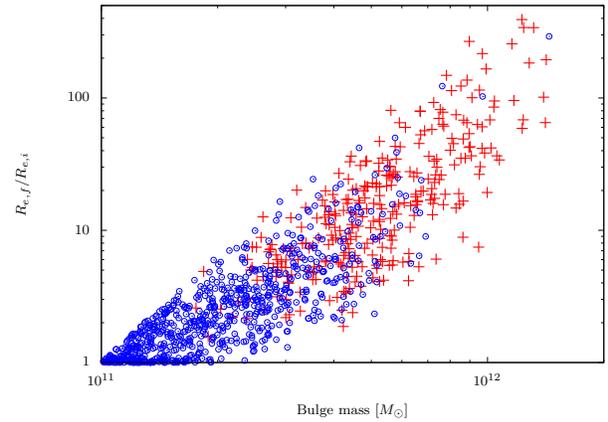}}
  \caption{
  The size growth factor $R_{e,f} / R_{e,i} \equiv (M_f/M_i)^{\alpha}$ of our sample BCGs (red crosses) and normal ETGs (blue open circles) derived by the Millennium Simulation Database, 
  where we assume $\alpha = 2.7$ which is the most efficient case in our results.
  Horizontal axis shows the bulge mass at $z=0$.
  Vertical axis shows $R_{e,f} / R_{e,i}$.
  } 
  \label{size_growth2}
\end{figure}

\begin{figure}
  \resizebox{84mm}{!}{\input{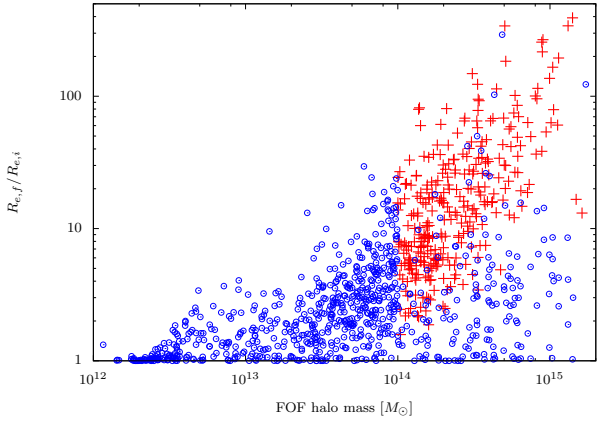}}
  \caption{
  The same as Fig. \ref{size_growth2}, but for the horizontal axis that shows the FOF halo mass which the galaxies belong to at $z=0$.
  } 
  \label{size_growth4}
\end{figure}

In Fig. \ref{size_growth2}, we show the individual size growth factor for all sample galaxies as a function of the bulge mass at $z=0$.
Fig. \ref{size_growth4} is the same as Fig. \ref{size_growth2} but as a function of the FOF halo mass to which the each sample galaxy belongs at $z=0$.
In this figure, we assume $\alpha = 2.7$.
We note that there are a considerable number of the galaxies which have a small size growth factor from $z=2.07$ to $z=0$ from dry minor mergers (Fig. \ref{size_growth2}).
These galaxies reside in dark matter haloes which have a wide range of mass from galaxy-size haloes to cluster-size ones (Fig. \ref{size_growth4}), and almost all of them are normal ETGs.
In addition, they also do not increase their size by dry major mergers.
In fact, in the galaxies which have a size growth factor less than two, $\sim 80\%$ are minor merger-dominated galaxies.
Thus, these galaxies are expected to keep their size from $z\sim 2$.
Such galaxies in cluster-size haloes may correspond to nearby compact massive ETGs which \citet{2010ApJ...712..226V} claim to find in nearby X-ray-selected clusters.

\subsection{Comparison with previous work}

Our simulation results show a high size growth efficiency and decrease in the velocity dispersion even in our lowest growth efficiency case with simultaneous minor mergers, compared with results from \citet{2009ApJ...706L..86N}.
This disagreement may be due to the initial conditions of the simulations.
In this section, we examine the differences between our simulations and those performed by \citet{2009ApJ...706L..86N} .
\citet{2009ApJ...706L..86N} construct initial galaxy models which have a higher dark matter fraction within the stellar half mass radius than our models.
The correct dark matter fraction with the half mass radii is uncertain for the compact massive ETGs at high-z. 
\citet{2010ApJ...721.1755S} show that massive compact high-z galaxies are strongly baryon dominated in their inner regions through cosmological simulations.
The semi-analytic model of \citet{2011arXiv1105.6043S} also predict that the dark matter fraction of compact massive ETGs become smaller in high-z.
\citet{2009ApJ...706L..86N} also assume a steeper inner slope, $\rho \propto r^{-1.5}$, than the Hernquist profile ($\rho \propto r^{-1}$) for the stellar component.
Their initial stellar model may preserve the inner structure during dry mergers, and may bind the satellite galaxies in the inner region.
These lead to suppression of significant size growth and a lower decrease in the velocity dispersion.
\citet{2009ApJ...706L..86N} assume more compact satellite galaxies than model B of our simulation.
This is because they scale the satellite galaxies based on the fundamental plane derived by the SLACS survey.
For model B in our simulations, we chose to simply scale down the primary galaxy (model A) with the same average density.
This is because we assume that formation epoch of model B is the same as that of model A, that is z = 2.0, and that these models are expected to have the same average density.
This assumption may mean that in the dry minor merger simulations, the effect of dynamical friction heating may be more effective than in our results.

\section{Conclusions}

We investigate the effects of sequential dry minor mergers on the size evolution of the compact massive ETGs.
From an analysis of the Millennium Simulation Database, we show that such minor mergers are highly likely in the hierarchical structure formation.
We perform N-body simulations of sequential minor mergers with parabolic and head-on orbits, including a dark matter and stellar component, 
and compare the properties of the merger remnants with that from simultaneous minor mergers.
We show that the sequential minor mergers are the most efficient for size growth and in decreasing the velocity dispersion of the stellar system.
The evolution of stellar size and velocity dispersion of the galaxy in the sequential minor mergers with compact satellites agrees with recent observations.
Furthermore, we construct the merger histories of candidates of the high-z compact massive ETGs using the Millennium Simulation Database, and model the size growth of the galaxies via dry minor mergers.
We can reproduce the mean size growth factor between $z=2$ and $z=0$ if we adopt the most efficient size growth efficiency $\alpha=2.7$ obtained in the case of sequential minor mergers in our simulations.
Because these minor mergers are more common for the galaxies in more massive FOF haloes (Fig. \ref{minor_dominated_fraction}), our results suggest that for ETGs in more massive dark matter haloes, the size growth by the dry minor mergers is still more effective.
However, we note that our numerical result is valid for merger histories with typical mass ratios between 1/20 and 1/10 and with parabolic and head-on orbits.
In addition, our most efficient size growth efficiency is likely to an upper limit since it is based on the simulations in which the satellites have smaller mass ratios and are on radial orbits.
In future studies, we will study various cases of mass ratios and merger orbits for a more robust prediction.

In this study, we did not focus the ``tightness'' of the local scaling relations of ETGs.
\citet{2009ApJ...706L..86N} note that it is difficult to explain such an exact relation through the mass and size growth of the compact massive ETGs at high-z with dry mergers. 
We will investigate the strength of the correlation during the cosmological merger histories in our future studies.

\section*{Acknowledgments}

We thank Volker Springel for making GADGET-2 publicly available.
We thank M. Fujimoto, E. Tasker and C. Nipoti for useful suggestions.
Numerical computations were carried out on Cray XT4 at Center for Computational Astrophysics, 
CfCA, of National Astronomical Observatory of Japan.
The Millennium Simulation databases used in this paper and the web application providing online access to them were constructed as part of the activities of the German Astrophysical Virtual Observatory.
This work was supported by Grant-in-Aid for Specially Promoted Research 20001003.

\bibliographystyle{mn}
\bibliography{mn-jour,ref}

\clearpage

\clearpage

\clearpage

\begin{figure}
    \begin{minipage}{84mm}
        \begin{center}
            \begin{tabular}{cc}
                \hspace{-10mm}
                \resizebox{55mm}{!}{\input{./picture/Rlog_Vdisplinear_1_2.tex}} &
                \hspace{-18mm}
                \resizebox{55mm}{!}{\input{./picture/Rlog_Vdisplinear_2_2.tex}} \\
                \hspace{-10mm}
                \resizebox{55mm}{!}{\input{./picture/anisotropy4_2.tex}} &
                \hspace{-18mm}
                \resizebox{55mm}{!}{\input{./picture/anisotropy5_2.tex}} \\
            \end{tabular}
            \caption{Top-left: Line-of-sight velocity dispersion profiles of the remnant stellar systems, 1A10Bsm(dotted), 1A10Bsq(thick solid), and initial model A (thin solid).
            Top-rightt: The same as the top-left panel, but for 1A10Csm(dotted), 1A10Csq(thick solid), and initial model A (thin solid).
            Bottom: anisotropy parameter $\beta$ for the merger remnants in the top panels.}
            \label{fig:Rlog_Vdisplinear}
        \end{center}
    \end{minipage}
\end{figure}

\clearpage

\newpage

\bsp

\label{lastpage}

\end{document}